\documentclass[fleqn,usenatbib]{mnras}

\usepackage[T1]{fontenc}
\usepackage{newtxtext,newtxmath}
\usepackage{orcidlink}

\usepackage{graphicx}
\usepackage{amsmath}
\newif\ifshowalttext
\showalttextfalse
\newcommand{\figurealttext}[1]{\ifshowalttext\par\smallskip\noindent{\footnotesize\textbf{Alt text:} #1\par}\fi}

\makeatletter
\patchcmd{\caption}{\refstepcounter}{\H@refstepcounter}{}{}
\makeatother

\title[Fast dynamical microlensing]{Microcaustics: Fast GPU-Accelerated Dynamical Microlensing of Time-Dependent Extended Sources}

\author[J. Fagin et al.]{
Joshua~Fagin\orcidlink{0000-0001-8723-6136},$^{1,2,3,4}$\thanks{\protect\hypertarget{Hfootnote.1}{}E-mail: joshua.fagin@stonybrook.edu}
Connor~Stone\orcidlink{0000-0002-9086-6398},$^{5,6,7,8,9}$
James~Hung-Hsu~Chan\orcidlink{0000-0001-8797-725X},$^{3,4}$
\newauthor
Sophia~Miskiewicz\orcidlink{0000-0003-0631-9701},$^{2,3,4}$
Henry~Best\orcidlink{0009-0009-6932-6379},$^1$
Paras~Sharma\orcidlink{0000-0002-6909-205X},$^1$
Matthew~O'Dowd\orcidlink{0009-0000-4476-5003},$^{2,3,4}$
\newauthor
Laurence~Perreault-Levasseur\orcidlink{0000-0003-3544-3939},$^{7,8,9,10}$
Yashar~Hezaveh\orcidlink{0000-0002-8669-5733}$^{7,8,9,10}$
and Simon~Birrer\orcidlink{0000-0003-3195-5507}$^1$
\\
$^1$Department of Physics and Astronomy, Stony Brook University, Stony Brook, NY 11794, USA\\
$^2$The Graduate Center of the City University of New York, 365 Fifth Avenue, New York, NY 10016, USA\\
$^3$Department of Astrophysics, American Museum of Natural History, Central Park West and 79th Street, NY 10024-5192, USA\\
$^4$Department of Physics and Astronomy, Lehman College of the CUNY, Bronx, NY 10468, USA\\
$^5$Dunlap Institute, University of Toronto, Toronto, Canada\\
$^6$Canadian Data Intensive Astrophysics PLatform (CanDIAPL)\\
$^7$Ciela - Montreal Institute for Astrophysical Data Analysis and Machine Learning, Montr\'eal, Canada\\
$^8$Department of Physics, Université de Montréal 2900 Edouard Montpetit Blvd Montréal Québec H3T 1J4, Canada\\
$^9$Mila - Quebec Artificial Intelligence Institute, Montr\'eal, Canada\\
$^{10}$Trottier Space Institute, Montr\'eal, Canada
}

\date{Accepted XXX. Received YYY; in original form ZZZ}
\pubyear{\the\year{}}

\begin{document}
\label{firstpage}
\pagerange{\pageref{firstpage}--\pageref{lastpage}}
\maketitle

\begin{abstract}
Gravitational microlensing probes unresolved quasar accretion disc structure and compact objects in lens galaxies and is an important systematic in strong-lens time-delay cosmography. Most analyses use static microlensing map simulations because evolving individual microlenses and caustics is computationally prohibitive. We present \textsc{Microcaustics}, a GPU-accelerated Python package built on PyTorch for dynamical microlensing of time-dependent extended sources in strong gravitational lenses. It combines a local-exact/far-field deflection solver, source-directed lens-plane scouting with temporal reuse, and biquadratic inverse polygon mapping (IPM). GPU batching and fused Triton rasterization accelerate the calculation, while binary caustic labels track source-centred crossings. Across three simulated quadruply lensed quasar systems, our package generates a 10-year, 147-frame multiband light curve in an amortized median of $0.68\,\mathrm{s}$ ($0.88\,\mathrm{s}$ with labels) on a consumer GPU with a median light-curve RMSE of $0.35\,\mathrm{mmag}$, while recovering every reference source-centred caustic crossing. At comparable map accuracy, the batched throughput is $780$ times higher than that of a recent CUDA IPM implementation and extrapolates to a speedup of millions relative to classic GPU-D. We demonstrate that our IPM converges as $N_{\rm ray}^{-0.84}$, approaching the idealized $N_{\rm ray}^{-1}$ convergence rate of piecewise-affine IPM and improving upon the $N_{\rm ray}^{-1/2}$ inverse ray-shooting rate. We also introduce a GPU-resident general relativistic and reverberating quasar source model and demonstrate our method with an evolving supernova source. These advances make fully dynamical stellar fields practical for Bayesian and simulation-based inference.
\end{abstract}

\begin{keywords}
gravitational lensing: micro -- gravitational lensing: strong -- methods: numerical -- software: development -- accretion, accretion discs -- quasars: general
\end{keywords}

\section{Introduction}
\label{sec:introduction}

Quasars are luminous active galactic nuclei (AGN) powered by accretion onto supermassive black holes \citep{Salpeter64,Zeldovich64}. Their ultraviolet and optical continuum originates on scales too small to resolve directly at cosmological distances. Strong gravitational lensing produces multiple macroimages of the same quasar, with different magnifications and arrival times \citep{Shajib2024StrongLensing}. Stars and other compact objects in the foreground galaxy further split each macroimage into unresolved microimages. Their combined magnification depends on the source's position, size, and spatial structure. Critical curves are lens-plane locations where the lens-mapping Jacobian determinant vanishes. They map to source-plane caustics, where point-source magnification formally diverges and across which the microimage configuration changes. Compact sources can undergo sharp magnification changes near these caustics, whereas extended sources average over the same structure. Microlensing therefore probes quasar structure far below the angular resolution of a telescope \citep{SchneiderEhlersFalco1992,VernardosReview2024}.

\begin{figure*}
    \centering
    \includegraphics[width=\linewidth]{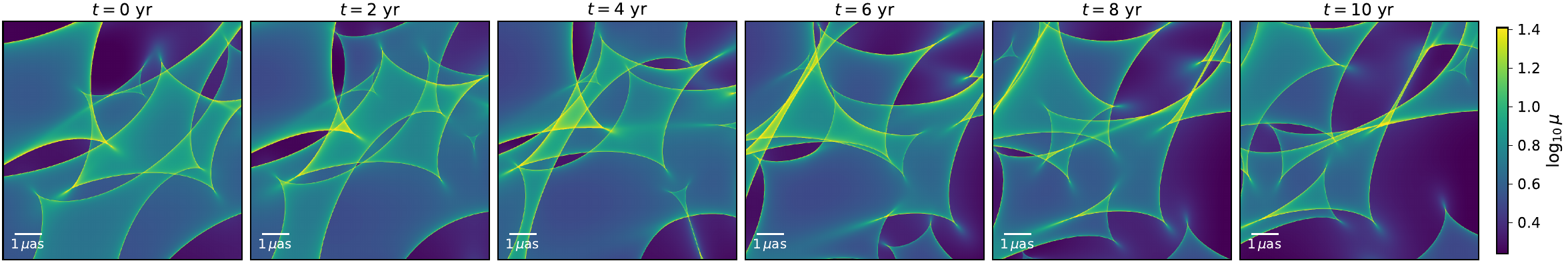}
    \caption{
Motivation for dynamical magnification maps using the Q2237+0305 image B parameters $\kappa=\gamma=0.391$, $s=0$, and $\phi_\gamma=141.73^\circ$. The stellar realization contains 3805 microlenses. The six panels show one macroimage at $t=0,2,4,6,8,$ and $10$ yr using a fixed $9.382\,\mu\mathrm{as}$ square source-plane window, $1024\times1024$ pixel grid, and shared $\log_{10}\mu$ normalization.
        }
    \figurealttext{Six magnification maps at two-year intervals show bright caustic ridges shifting, merging, and changing shape as the stars move. All panels use the same source-plane window and colour scale.}
\label{fig:dynamic_map_motivation}
\end{figure*}

Most quasar microlensing calculations convolve a static magnification map with a source profile and extract light curves along trajectories through that map. This freezes the caustic network. Stellar proper motions instead reshape the network and can produce shorter, steeper, and more frequent events than translation across a fixed pattern \citep{KundicWambsganss1993,WambsganssKundic1995}. Fig.~\ref{fig:dynamic_map_motivation} illustrates this evolution. Recomputing maps and consistently labelling caustics at every epoch has historically been prohibitively expensive for large simulation sets. \citet{Poindexter10Methods,Poindexter10} included moving stars in Bayesian analyses of Q2237+0305, with complete inference calculations requiring CPU-years of computation. Making such dynamical calculations inexpensive is the main goal of this work.

The scientific applications extend beyond measuring a single source size. Wavelength-dependent microlensing constrains the disc temperature profile, inclination, and spatial structure \citep{Poindexter08,Poindexter10,Blackburne15,Munoz16,Morgan18,Cornachione_2020,Best_2024,VernardosReview2024}. Measurements often find ultraviolet--optical discs several times larger than luminosity-based thin-disc predictions, with ensemble studies favouring shallower effective temperature profiles from microlensing \citep{Morgan2010,Cornachione_2020} and reverberation mapping \citep{Mudd2018,Guo_2022,Jha_2022}. A high-magnification event (HME), produced by a caustic crossing or close approach, can scan different disc radii and relativistically brightened regions \citep{Best_2024}. Forecasting these events can trigger high-cadence multiwavelength or spectroscopic follow-up \citep{fagin2024predictinghighmagnificationevents}.

The source itself also evolves. In lamppost reprocessing models, coronal variability illuminates the disc with position-dependent light-travel delays \citep{Cackett2007}. Microlensing then weights regions with different temperatures, variability amplitudes, and reverberation delays \citep{Vernardos_2019,Millon2022SMBHB,Best_2024}. It can alter both the observed variability and the inferred delay between macroimages \citep{TieKochanek2018,Chan2021MicrolensingTimeDelay}. Those delays constrain the cosmic expansion rate when combined with deflector and line-of-sight models \citep{Refsdal1964,TreuMarshall2016,Birrer2024TimeDelayCosmography}, making microlensing a source of cosmographic systematics as well as a probe of disc structure. Evolving transients, including strongly lensed supernovae, likewise require time-dependent source profiles \citep{Huber_2021}.

Microlensing also constrains the foreground lens. Compact-source flux anomalies complicate the use of macroimage flux ratios to measure dark-matter substructure \citep{MaoSchneider1998,DalalKochanek2002,SchechterWambsganss2002,Nierenberg2024JWST,Gilman2026JWST}. Multi-epoch and multiwavelength observations help separate stellar microlensing from more nearly static substructure \citep{VernardosReview2024}. Magnification statistics and event time-scales also depend on the microlens mass spectrum, allowing ensembles to constrain the mean stellar mass and, with additional information, extragalactic stellar initial mass functions \citep{JimenezVicenteMediavilla2019}. Related applications include compact dark matter \citep{Mediavilla2017PBH}, wandering intermediate-mass black holes \citep{WuHo2025}, and extragalactic planets \citep{DaiGuerras2018,Tuntsov2024Planets}. These inferences benefit from forward models that vary source structure, velocities, and microlens masses together.

Large precomputed maps are effective for sampling many trajectories through a static field, but a dynamical library adds time as another dimension. The more than 70,000 static GERLUMPH maps already occupy roughly 25~TB \citep{Vernardos2014,VernardosFluke2014Cloud}. Reusing a few realizations limits training diversity, while on-demand simulation can redraw the stars and source without storing a mostly unused map. A precomputed map also fixes the smallest spatial scales available to subsequent convolution and interpolation, which can limit studies of inner-disc emission and early supernova photospheres unless finer maps are generated. Population studies also commonly adopt a single microlens mass and a circular, often Gaussian source \citep{Vernardos_2019,Neira2020,Neira_2025,fagin2024predictinghighmagnificationevents}. Such profiles can reproduce broad statistics at fixed half-light radius \citep{Mortonson2005}, but discard structure that matters during caustic crossings. Intrinsic variability is often removed approximately through time-delay-corrected difference light curves or modelled as a nuisance process \citep{SluseTewes2014,TieKochanek2018LightCurves,Paic2022}. A resolved evolving source instead allows intrinsic variability and microlensing to be calculated together \citep{best2024amoebaagnmodeloptical}.

Inverse ray shooting (IRS) constructs maps by accumulating deflected rays in source pixels \citep{Kayser1986,SchneiderWeiss1987}, whereas inverse polygon mapping (IPM) distributes mapped cell areas among those pixels \citep{Mediavilla2006,Mediavilla2011}. Tree and direct GPU methods accelerated IRS \citep{Wambsganss1990,Bate2010,Thompson2010}, with GPU-D underpinning GERLUMPH \citep{Vernardos2014}. Later implementations use mesh-based Poisson solvers \citep{Shalyapin2021PIP}, interpolated far fields \citep{Zheng2022GPU}, or fast multipole methods \citep{GreengardRokhlin1987,JimenezVicente2022}. The open-source GPU implementation of \citet{Weisenbach2025Maps} combines the fast multipole method with IPM, with a companion code for critical curves and caustics \citep{Weisenbach2025}. These methods emphasize accurate individual maps and caustic calculations. Together with adaptive spatial selection \citep{Meena2022AdaptiveBoundary}, they provide the basis for an end-to-end GPU calculation optimized for evolving stellar fields and extended sources.

The Vera C. Rubin Observatory Legacy Survey of Space and Time (LSST) makes this scale of simulation increasingly important. Forecasts predict more than 8000 lensed quasars, including roughly 3000 with well-measured time delays \citep{OguriMarshall2010}. LSST will monitor them in six near-UV to near-IR bands over ten years \citep{LSST}, with dozens of HMEs expected annually across the monitored lens population \citep{Neira2020,Neira_2025}, alongside lensed supernovae and other transients \citep{Sheng_2022,Arendse_2024}. Machine-learning methods for parameter inference, light-curve reconstruction, and HME prediction can require $\gtrsim10^6$ diverse multiband training curves \citep{Vernardos_2019,Fagin_2024,fagin2024predictinghighmagnificationevents,Fagin2025Reprocessing}. Bayesian Monte Carlo analyses likewise require many simulated trajectories to marginalize over lens, source, and velocity parameters \citep{Kochanek_2004,Poindexter10,Morgan_2012,Blackburne15,Cornachione_2020}.

Here we introduce \textsc{Microcaustics}, a GPU-accelerated Python framework for dynamical microlensing of time-dependent extended sources. It combines a local-exact/far-field deflection solver, source-directed lens-plane scouting, IPM, temporal reuse, and batched source evaluation in PyTorch and Triton \citep{Pytorch,Tillet2019Triton}. The simulator returns multiband light curves together with critical curves, caustics, and crossing labels. We compare its accuracy and performance with high-resolution references and existing GPU implementations. We also introduce a GPU-resident relativistic, reverberating quasar disc with an adjustable temperature slope and demonstrate an expanding supernova source. The code is publicly available at \url{https://github.com/JFagin/microcaustics}.

Sections~\ref{sec:simulator_formulation} and \ref{sec:numerical_method} define the physical model and numerical method, Section~\ref{sec:validation_results} presents the benchmarks, and Section~\ref{sec:source_models} describes the source models and end-to-end examples. We discuss applications and limitations in Section~\ref{sec:discussion} and present our conclusions in Section~\ref{sec:conclusions}. Throughout, we assume a flat $\Lambda$CDM cosmology with $H_0=70\,{\rm km\,s^{-1}\,Mpc^{-1}}$ and $\Omega_{\rm m}=0.3$, and report AB magnitudes.

\section{Dynamical Microlensing Model and Observables}
\label{sec:simulator_formulation}

\textsc{Microcaustics} separates the slowly varying galaxy-scale lens from the stellar-scale perturbations around each macroimage. The former determines the local convergence, shear, parity, and macroimage time delay, and the latter is evolved explicitly and coupled to a time-dependent extended source. This section defines that physical calculation and its observables. The accelerated evaluation of the same equations is described in Section~\ref{sec:numerical_method}, and the source models are specified in Section~\ref{sec:source_models}. The benchmark sample comprises the four macroimages of each of Q2237+0305, RX~J1131$-$1231, and HE~0435$-$1223, each simulated as a 10-year, microlensing-only light curve at 25-day intervals (147 frames). The mass and velocity prescriptions below define their fiducial calculations, and their adopted lens and source parameters are collected in Section~\ref{subsec:benchmark_sample}. The simulator also accepts user-supplied microlens masses, positions, and velocities.

\subsection{Local macroimage and microlens model}
\label{subsec:macro_lens}

We describe a small region around one macroimage using angular image-plane coordinates $\boldsymbol{\theta}=(\theta_1,\theta_2)$ and source-plane coordinates $\boldsymbol{\beta}=(\beta_1,\beta_2)$, both measured relative to the corresponding macroimage solution. A constant deflection is absorbed into the choice of origin. Over the stellar field, the remaining smooth galaxy deflection is well approximated by a local linear field described by convergence and shear, while the compact component is represented by individual point masses \citep{SchneiderEhlersFalco1992,VernardosReview2024}. The local lens equation is
\begin{equation}
\begin{split}
 \boldsymbol{\beta}(\boldsymbol{\theta},t)
 ={}& \mathbf{A}_{\rm s}\boldsymbol{\theta}
 -\sum_{k=1}^{N_\star}\theta_{{\rm E},k}^{\,2}
 \frac{\boldsymbol{\theta}-\boldsymbol{\theta}_k(t)}
 {\left|\boldsymbol{\theta}-\boldsymbol{\theta}_k(t)\right|^2},\\
 \mathbf{A}_{\rm s}={}&
 \begin{pmatrix}
 1-\kappa_{\rm s}-\gamma_1 & -\gamma_2\\
 -\gamma_2 & 1-\kappa_{\rm s}+\gamma_1
 \end{pmatrix}.
\end{split}
\label{eq:local_microlens}
\end{equation}
Here $\kappa_{\rm s}$ is the smoothly distributed convergence, $(\gamma_1,\gamma_2)$ are the shear components, and $\boldsymbol{\theta}_k(t)$ and $\theta_{{\rm E},k}$ are the position and angular Einstein radius of microlens $k$ (see equation~\eqref{eq:einstein_radius}). Differentiating equation~\eqref{eq:local_microlens} away from the microlens positions gives

\begin{equation}
 \mathbf J\equiv\frac{\partial\boldsymbol\beta}{\partial\boldsymbol\theta}
 =\mathbf A_{\rm s}-\sum_k\theta_{{\rm E},k}^2
 \left[\frac{\mathbf I}{d_k^2}
 -\frac{2\boldsymbol d_k\boldsymbol d_k^{\mathsf T}}{d_k^4}\right],
 \label{eq:local_lens_jacobian}
\end{equation}
where $\boldsymbol d_k=\boldsymbol\theta-\boldsymbol\theta_k(t)$, $d_k=|\boldsymbol d_k|$, and $\mathbf I$ is the identity matrix.

We write the total convergence as $\kappa=\kappa_{\rm s}+\kappa_\star$ and define the smooth-matter fraction $s=\kappa_{\rm s}/\kappa$, so that $\kappa_\star=(1-s)\kappa$. With
\begin{equation}
\begin{gathered}
 \gamma=(\gamma_1^2+\gamma_2^2)^{1/2},\qquad
 \phi_\gamma=\frac{1}{2}\operatorname{atan2}(\gamma_2,\gamma_1),\\
 \mu_{\rm macro}=\frac{1}{(1-\kappa)^2-\gamma^2},
\end{gathered}
\label{eq:macro_local_quantities}
\end{equation}
the sign of $\mu_{\rm macro}$ gives the macroimage parity. Positive-parity images are minima or maxima of the arrival-time surface, whereas negative-parity images are saddles \citep{SchneiderEhlersFalco1992,Shajib2024StrongLensing}.

For known lens systems, these local quantities can be supplied from an observational macro model. Alternatively, the simulator can solve a complete user-specified macro model, calculate its macroimage positions and relative arrival times, and evaluate the Jacobian at each image to obtain $\kappa$, $\gamma_1$, $\gamma_2$, and parity, as demonstrated in Section~\ref{subsec:hst_lsst}. Each macroimage then defines an independent realization of equation~\eqref{eq:local_microlens}.

\subsection{Dynamical microlens population}
\label{subsec:microlens_population}

For lens and source angular-diameter distances $D_{\rm l}$ and $D_{\rm s}$ and lens--source distance $D_{\rm ls}$, the angular Einstein radius corresponding to the mean microlens mass is

\begin{equation}
 \theta_{\rm E}(\langle M\rangle)
 = \sqrt{
 \frac{4G\langle M\rangle}{c^2}
 \frac{D_{\rm ls}}{D_{\rm l}D_{\rm s}}}, \;\;\;\;
 \theta_{{\rm E},k}
 =
 \theta_{\rm E}(\langle M\rangle)
 \sqrt{\frac{M_k}{\langle M\rangle}} .
\label{eq:einstein_radius}
\end{equation}
We draw microlens masses from a truncated power law $p(M)\propto M^{-\alpha}$ for $M_{\min}\leq M\leq M_{\max}$, with the Salpeter slope $\alpha=2.35$ \citep{Salpeter1955,Kochanek_2004}. The distribution is specified by its mean and dynamic range. Every benchmark uses $\langle M\rangle=0.3\,M_\odot$ and $M_{\max}/M_{\min}=100$.

Initial microlens positions are uniform within a circular image-plane field. For a diameter $L_\star$ in Einstein-angle units, the population is $N_\star=\operatorname{round}(\kappa_\star L_\star^2/4)$. We size the field using the large-deflection tail criterion \citep{KatzBalbusPaczynski1986,Weisenbach2025Maps}, then circumscribe the protected macroimage region with a circle, apply a safety factor $s_{\rm safe}=1.5$, and pad for coherent motion. With source-field side length $\ell_{\rm src}=L_{\rm source}/\theta_{\rm E}$, mass-moment factor $C_M=\langle M^2\rangle/\langle M\rangle^2$, and nominal tail probability $f_{\rm loss}=0.01$, the shear-aligned half-widths and final stellar diameter are
\begin{equation}
 \begin{aligned}
 b_{\rm src}&=\frac{\ell_{\rm src}}{2}
       +\sqrt{\frac{\kappa_\star C_M}{f_{\rm loss}}},&
 b_1&=\frac{b_{\rm src}}{|1-\kappa-\gamma|},\\
 b_2&=\frac{b_{\rm src}}{|1-\kappa+\gamma|},&
 L_\star&=2s_{\rm safe}\sqrt{b_1^2+b_2^2}
             +|\boldsymbol\omega_{\rm bulk}|T.
 \end{aligned}
 \label{eq:stellar_field_diameter}
\end{equation}
Here $b_{\rm src}$, $b_1$, $b_2$, and $L_\star$ are in Einstein-angle units, $\boldsymbol\omega_{\rm bulk}$ is the coherent image-plane motion in those units per observer day, and $T$ is the duration in observer days. The final term enlarges the stellar field for longer simulations so that the protected source region remains covered along the complete coherent track. The reported physical angular diameter is $L_{\rm lens} = \theta_{\rm E}L_\star$.

We retain the full two-dimensional velocity vectors rather than assigning a single scalar track speed. Let $\boldsymbol{v}_{\rm l}$ and $\boldsymbol{v}_{\rm s}$ be the proper transverse peculiar velocities of the lens and source in their local rest frames, and let $\boldsymbol{v}_{\rm o}$ be the observer velocity projected onto the tangent plane at the lens coordinates. In units of mean-mass Einstein radii per unit observer time, the coherent motion applied to the microlens field is
\begin{equation}
 \boldsymbol{\omega}_{\rm bulk} =
 \frac{\boldsymbol{v}_{\rm l}}
 {(1+z_{\rm l})D_{\rm l}\theta_{\rm E}}
 -
 \frac{\boldsymbol{v}_{\rm s}}
 {(1+z_{\rm s})D_{\rm s}\theta_{\rm E}}
 -
 \frac{D_{\rm ls}\boldsymbol{v}_{\rm o}}
 {(1+z_{\rm l})D_{\rm l}D_{\rm s}\theta_{\rm E}} .
\label{eq:bulk_angular_velocity}
\end{equation}
We implement coherent bulk motion in a source-centred frame. Rather than moving the source through the magnification pattern, we hold the source fixed and translate the local lens model by $\boldsymbol d_{\rm bulk}(t)=\theta_{\rm E}\boldsymbol\omega_{\rm bulk}t$. This translation includes both the microlenses and the smooth lens component. Accounting for the latter adds the uniform term $(\mathbf I-\mathbf A_{\rm s})\boldsymbol d_{\rm bulk}$ to the lens mapping. With internal stellar motions switched off, this gives
\begin{equation}
 \boldsymbol\beta_t(\boldsymbol\theta)
 =\boldsymbol\beta_0(\boldsymbol\theta-\boldsymbol d_{\rm bulk})
 +\boldsymbol d_{\rm bulk}.
\label{eq:bulk_mapping_offset}
\end{equation}
The equivalent source moves through the stationary pattern in the opposite direction, with effective velocity
\begin{equation}
 \boldsymbol{v}_{\rm eff}^{({\rm s})}
 \equiv-D_{\rm s}\theta_{\rm E}\boldsymbol{\omega}_{\rm bulk}.
\label{eq:source_plane_effective_velocity}
\end{equation}
Substituting equation~\eqref{eq:bulk_angular_velocity} recovers the standard geometrical projection \citep{Kayser1986,MosqueraKochanek2011}. Writing the line-of-sight unit vector as $\hat{\boldsymbol{n}}$, the code evaluates $\boldsymbol{v}_{\rm o}= [\mathbf{I}-\hat{\boldsymbol{n}}\hat{\boldsymbol{n}}^{\mathsf T}] \boldsymbol{v}_{\rm CMB}$ and resolves it into local $x$ and $y$ components. We use the measured Solar-system cosmic microwave background (CMB) dipole velocity $|\boldsymbol{v}_{\rm CMB}|=369.82\ {\rm km\,s^{-1}}$ toward Galactic coordinates $(l,b)=(264.021^\circ,48.253^\circ)$ \citep{Planck2020Overview}. Thus the observer contribution has a fixed direction for a given lens rather than a newly drawn random orientation.

For $j\in\{\mathrm{lens},\mathrm{source}\}$, the Cartesian velocity components indexed by $a\in\{x,y\}$ are drawn independently with the same dispersion,
\begin{equation}
 v_{j,a}\sim\mathcal{N}\!\left[0,\sigma_{\rm pec}^{2}(z_j)\right],\qquad
 \sigma_{\rm pec}(z)=\frac{\sigma_0}{\sqrt{1+z}}\frac{f(z)}{f(0)}.
\label{eq:peculiar_velocity_dispersion}
\end{equation}
Here $f=d\ln D_+/d\ln a$ is the linear growth rate, with cosmological scale factor $a$, evaluated using the approximation of \citet{LightmanSchechter1990,Lahav1991} as written in equation (5) of \citet{Hamilton2001}. The redshift scaling and normalization $\sigma_0=235\ {\rm km\,s^{-1}}$ follow \citet{Kochanek_2004}.

Finally, internal stellar motions are not absorbed into $\boldsymbol{v}_{\rm eff}^{({\rm s})}$. Each internal tangent-plane velocity component is drawn as
\begin{equation}
\begin{split}
 \delta\omega_{k,a}
 &\sim\mathcal{N}(0,\sigma_{\omega,\star}^{2}),\qquad
 \sigma_{\omega,\star}
 =\frac{\sigma_\star}
 {(1+z_{\rm l})D_{\rm l}\theta_{\rm E}}.
\end{split}
\label{eq:stellar_motion}
\end{equation}
Here $\sigma_\star$ is the one-dimensional stellar velocity dispersion of the lens. We adopt $\sigma_\star=(170,323,222)\ {\rm km\,s^{-1}}$ for Q2237+0305, RX~J1131$-$1231, and HE~0435$-$1223, respectively \citep{Chen2019SHARP,Tuntsov2024Planets}. The Q2237 value is an isothermal-model-based fiducial \citep{vanDeVen2010}. Assuming isotropic stellar velocities, the root-mean-square two-dimensional stellar speed is $\sqrt{2}\sigma_\star$, but our stellar-kinematics calculation draws an independent vector for every star rather than replacing the evolving pattern by one bulk vector of that magnitude, as is commonly done for static-map light-curve generation \citep[e.g.,][]{Neira2020, Neira_2025}. Microlens trajectories are advanced with angular velocity $\theta_{\rm E}\delta\boldsymbol\omega_k$ and specularly reflected within the circular stellar aperture to maintain a stationary finite population. The aperture's safety and motion padding place this boundary far from the protected source-reaching region, so the reflected trajectories should have negligible influence on the central calculation. Reflection prevents microlenses from diffusing out of the finite aperture, which would otherwise change the stellar centre of mass and the effective local macro parameters. The coherent bulk displacement is then applied to the entire field. The resulting differential motion changes both the locations and topology of the caustic network and is not, in general, equivalent to translating a source through one static map \citep{WambsganssKundic1995}. The relative importance of this deformation is set in part by the ratio of the root-mean-square internal stellar motion to the coherent effective motion, $\sqrt{2}\sigma_{\omega,\star}/|\boldsymbol{\omega}_{\rm bulk}|$. Larger ratios produce stronger changes in the caustic pattern relative to its bulk translation. Across the 12 benchmark macroimages, this ratio has a median of $0.80$ and an observed 16th--84th percentile range of $0.65$--$1.2$.

\subsection{Time-dependent finite-source observables}
\label{subsec:finite_source_observables}

Let $I_{\nu,b}(\boldsymbol{\beta},t)$ be the observer-frame specific surface brightness of an arbitrary source in band $b$. The point-source magnification at source position $\boldsymbol{\beta}$ is the sum over all of its microimages,
\begin{equation}
 \mu_i(\boldsymbol{\beta},t)
 =
 \sum_q
 \frac{1}
 {\left|\det\mathbf{J}_i(\boldsymbol{\theta}_q,t)\right|},
\label{eq:point_magnification}
\end{equation}
where $i$ identifies the macroimage, $q$ runs over all microimage solutions $\boldsymbol{\beta}(\boldsymbol{\theta}_q,t)=\boldsymbol{\beta}$ within that macroimage, and the Jacobian is defined in equation~\eqref{eq:local_lens_jacobian}. This absolute magnification includes the macro model. Its large-area or ensemble mean approaches $|\mu_{{\rm macro},i}|$, but an individual small map need not have this mean.

Gravitational lensing preserves surface brightness \citep{SchneiderEhlersFalco1992}. At observer time $t$, the lensed band flux is
\begin{equation}
 F_{\nu,i,b}^{\,\rm L}(t)=
 \int_{\Omega_\beta}\mu_i(\boldsymbol{\beta},t)
 I_{\nu,b}(\boldsymbol{\beta},t-\Delta t_i)\,
 d^2\boldsymbol{\beta},
\label{eq:finite_source_flux}
\end{equation}
where $\Omega_\beta$ is the requested source-plane region and $\Delta t_i$ is the macroimage delay. Setting $\mu_i=1$ and $\Delta t_i=0$ recovers the unlensed flux at observer time $t$. Numerically, we sum over source pixels, combining each magnification map with the requested source images.

The macroimage delays are supplied or calculated from the macro model using the standard Fermat-potential differences \citep{Refsdal1964,TreuMarshall2016}. We set the reference image's delay to zero, with positive delays corresponding to later arrival.

Note that nothing in the preceding section requires a particular source morphology. The simulator needs only a surface-brightness image or image sequence on a known physical and angular scale. The general relativistic quasar, supernova, and generic source interfaces are described in Section~\ref{sec:source_models}.

\subsection{Critical curves and caustics}
\label{subsec:caustic_products}

The critical curves at time $t$ are the image-plane zero set of the lensing Jacobian determinant, and their mapping through the lens equation produces the source-plane caustics:
\begin{equation}
 \mathcal{C}_\theta(t)
 =\{\boldsymbol{\theta}:\det\mathbf{J}(\boldsymbol{\theta},t)=0\},
 \qquad
 \mathcal{C}_\beta(t)
 =\boldsymbol{\beta}\!\left[\mathcal{C}_\theta(t),t\right].
\label{eq:critical_caustic}
\end{equation}
Point-source magnification formally diverges on a caustic, whereas the integral in equation~\eqref{eq:finite_source_flux} remains finite for an extended source. Across a generic fold caustic, a pair of microimages is created or destroyed \citep{SchneiderEhlersFalco1992,VernardosReview2024}.

\section{Efficient GPU-Accelerated Dynamical Microlensing Simulation}
\label{sec:numerical_method}

\subsection{Numerical pipeline}
\label{subsec:pipeline}

The direct evaluation of equation~\eqref{eq:local_microlens} for $N_{\rm ray}$ image-plane positions and $N_\star$ microlenses requires $\mathcal{O}(N_{\rm ray}N_\star)$ pairwise interactions. A conventional IRS calculation incurs this cost over a field large enough to contain most of the rays that can map into the source plane, even though a compact source generally receives flux from only a small and irregular subset of that field. A standard dynamical calculation repeats the operation at every epoch.

In contrast, \textsc{Microcaustics} reduces the costs of many-microlens deflection, noncontributing lens-plane area, finite-cell integration, and repeated dynamical evaluation. Several epochs and independent light curves are processed together on the GPU. For each epoch, we update the stellar positions and construct the local-exact/far-field deflection approximation (Section~\ref{subsec:far_field_approx}). A coarse scout identifies lens-plane cells whose mapped footprints can contribute to the source (Section~\ref{subsec:lens_plane_scout}). Nearby epochs reuse this cell selection while recalculating the lens mapping with the current stellar positions (Section~\ref{subsec:temporal_scout_reuse}).

Only the selected region is traced at fine resolution. Within each cell, a biquadratic reconstruction supplies a finer virtual lattice without additional lens-equation evaluations, and IPM distributes the mapped areas among source pixels (Section~\ref{subsec:biquadratic_ipm}). We multiply the resulting magnification maps by the multiband source brightness and sum over pixels to obtain the light curves (Section~\ref{subsec:map_flux_batching}). When requested, a separate calculation evaluates the Jacobian in the selected regions, extracts critical curves, maps them into source-plane caustics, and assigns caustic-crossing labels (Section~\ref{subsec:anchor_gauge}).

\subsection{Local-exact/far-field deflection solver}
\label{subsec:far_field_approx}

\begin{figure*}
    \centering
    \includegraphics[width=\linewidth]{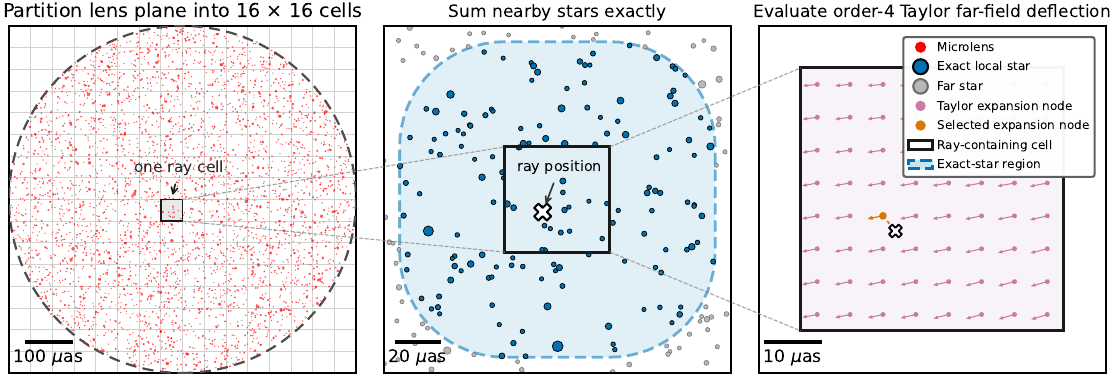}
    \caption{
Local-exact/far-field deflection solver diagram for a Q2237+0305 image-B realization. \emph{Left:} the circular microlens field is partitioned into $16\times16$ spatial cells. \emph{Middle:} for a ray in one cell, microlenses within the one cell width local neighbourhood are summed exactly, while the remaining microlenses define a smooth far field. \emph{Right:} fourth-order Taylor expansions about the internal nodes of the $8\times8$ lattice approximate the far-field deflection at each ray position. Grey dashed connectors indicate successive enlargement of the same ray-containing cell across the three panels.
    }
    \figurealttext{Three progressively enlarged diagrams show the circular stellar field, one partition cell with nearby stars, and the interpolation lattice inside that cell. Nearby deflections are evaluated directly and the more distant contribution is approximated.}
\label{fig:far_field_schematic}
\end{figure*}

Nearby microlenses can produce rapidly varying deflections, whereas the combined contribution from distant microlenses varies smoothly across a small region. We evaluate nearby stars individually and approximate the distant contribution with a polynomial that can be evaluated cheaply at many ray positions. This near/far separation is an established acceleration strategy. In particular, \citet{Zheng2022GPU} combine direct nearby deflections with local Taylor expansions of the distant contribution.

We partition the stellar field into spatial cells. For a ray at $\boldsymbol{\theta}$ in cell $c$, we approximate the point-mass deflection as
\begin{equation}
 \boldsymbol{\alpha}_{\star}(\boldsymbol{\theta})
 \simeq
 \sum_{j\in\mathcal{L}_c}
 \theta_{{\rm E},j}^{2}
 \frac{\boldsymbol{r}_j}{r_j^{2}}
 +
 \boldsymbol{\mathcal P}_{c,\nu(\boldsymbol{\theta})}
 (\boldsymbol{\theta}),
 \qquad
 \boldsymbol{r}_j
 \equiv
 \boldsymbol{\theta}-\boldsymbol{\theta}_j,
\label{eq:far_field_deflection}
\end{equation}
where $\mathcal L_c$ contains nearby microlenses and $\mathcal F_c$ contains the remaining, distant microlenses. Within each cell, a regular lattice of nodes supplies local polynomial expansions. The index $\nu(\boldsymbol{\theta})$ selects the node at the centre of the subcell containing the ray, and $\boldsymbol{\mathcal P}_{c,\nu}$ gives the corresponding far-field deflection.

Complex notation combines the two deflection components into a single expression. We write $z=\theta_x+i\theta_y$ and represent the deflection by $\alpha_x+i\alpha_y$. A point mass then contributes $\theta_{{\rm E},j}^{2}/(\bar z-\bar z_j)$, where the bar denotes complex conjugation. Expanding about a node $z_{c,\nu}$ uses the geometric series $(1+x)^{-1}=\sum_{n=0}^{\infty}(-x)^n$, with $x=(\bar z-\bar z_{c,\nu})/(\bar z_{c,\nu}-\bar z_j)$ and $|x|<1$. Summing over distant stars and retaining terms through order $p$ gives the nodal expansion
\begin{equation}
 \mathcal{P}_{c,\nu}(\bar z)=
 \sum_{j\in\mathcal{F}_c}
 \frac{\theta_{{\rm E},j}^{2}}{\overline{z}_{c,\nu}-\overline{z}_j}
 \sum_{n=0}^{p}
 \left[-\frac{\overline{z}-\overline{z}_{c,\nu}}
 {\overline{z}_{c,\nu}-\overline{z}_j}\right]^n,
\label{eq:far_field_polynomial}
\end{equation}
where $p=4$ in the adopted configuration. Its real and imaginary parts give the two physical deflection components, so $\boldsymbol{\mathcal P}_{c,\nu}=(\operatorname{Re}\mathcal P_{c,\nu},\operatorname{Im}\mathcal P_{c,\nu})$. Contributions to each coefficient are combined before tracing the rays. Each ray therefore evaluates a short polynomial rather than summing over every distant microlens.

Calculating these coefficients directly at every node would require repeatedly summing over the same distant stars. Instead, we first sum their contributions into an order-ten expansion about the cell centre. We then shift this expansion to each internal node and retain terms through fourth order. Terms above fourth order contribute to the lower-order coefficients when the expansion centre is shifted, so keeping them during this step improves accuracy. Each ray evaluates only the resulting fourth-order polynomial.

Figure~\ref{fig:far_field_schematic} illustrates the adopted configuration. A $16\times16$ Cartesian partition covers the circular stellar aperture, with cells outside the aperture containing no microlenses. A microlens enters the exact local sum when its Euclidean distance from the ray-containing cell is smaller than $R_{\rm exact}=\chi_{\rm near}\max(\Delta\theta_{c,x},\Delta\theta_{c,y})$, with $\chi_{\rm near}=1$. Each cell contains an $8\times8$ nodal lattice. Each ray evaluates the polynomial centred on its selected node using its displacement from that node. These nodes store polynomial coefficients rather than deflection values for interpolation. Fourth order was the lowest tested evaluation order that kept the far-field error small at the adopted accuracy target. Higher orders increased storage and arithmetic without a useful reduction in the final error.

The lensing Jacobian uses the same decomposition. We sum the local point-mass derivatives exactly, differentiate the far-field polynomial analytically, and add the smooth macro-lens contribution in equation~\eqref{eq:local_lens_jacobian}. This avoids finite differencing and keeps the deflection and Jacobian internally consistent.

The nearby-star selection and far-field coefficients are recalculated at every dynamical epoch. Several epochs are processed together on the GPU, and rays assigned to the same node share its polynomial coefficients. For the temporal benchmark references, we use a more conservative $32\times32$ nodal lattice and exact-neighbourhood factor $\chi_{\rm near}=1.5$. Section~\ref{subsec:far_field_approx_results} tests the far-field approximation against direct point-mass deflections.

\subsection{Source-directed lens-plane scouting}
\label{subsec:lens_plane_scout}

\begin{figure*}
    \centering
    \includegraphics[width=\linewidth]{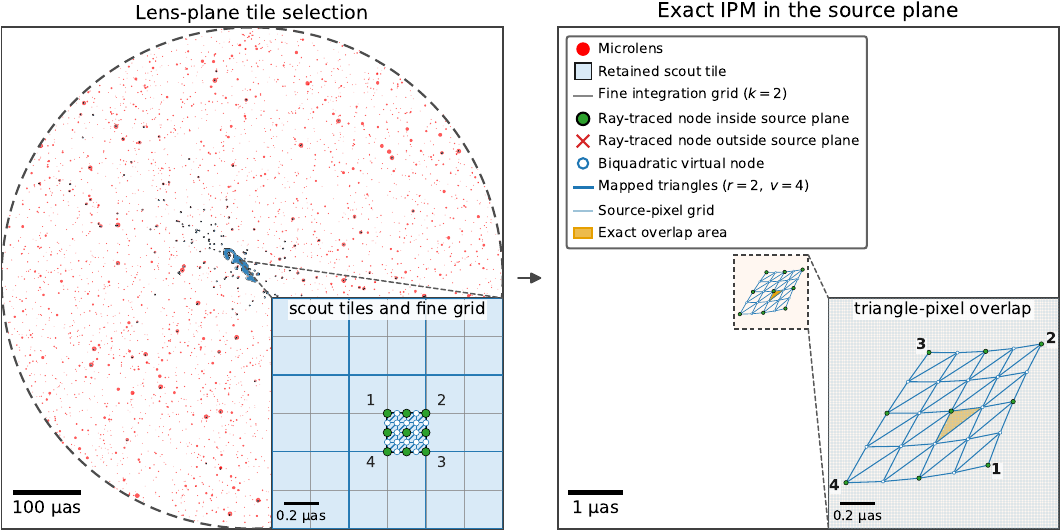}
    \caption{
Source-directed lens-plane scouting and biquadratic inverse polygon mapping for the standard $k=2$, $r=2$, and $v=4$ configuration. \emph{Left:} the nested scout grid is traced within the circular stellar aperture. A scout cell is selected when its mapped quadrilateral intersects the finite source. Here, the scouting retains only $0.26\%$ of the full-field cells. The inset shows its four fine-grid vertices and their mapped positions. \emph{Right:} a selected fine cell is evaluated on the $(r+1)\times(r+1)$ lattice of true ray-traced nodes. A biquadratic reconstruction supplies the virtual $v=4$ lattice, after which each virtual subcell is split into two mapped triangles whose areas are distributed according to their analytic overlap with the intersected source pixels, with an example triangle shown in orange.
    }
    \figurealttext{A two-part schematic follows selected lens-plane cells into the source plane. The scout identifies cells that intersect the source, then true and interpolated nodes divide each mapped cell into triangles for area-conserving deposition into source pixels.}
\label{fig:scout_ipm_schematic}
\end{figure*}

A compact source generally receives flux from only a sparse and irregular subset of the full image-plane field. Applying IPM or IRS to that entire field is therefore computationally inefficient. We first use a coarse scout to identify lens-plane cells whose mapped footprints intersect the requested source field, then perform the more expensive fine-grid calculation only in the selected region. The adaptive boundary method similarly refines image-plane regions that map near a compact source \citep{Meena2022AdaptiveBoundary}. Our scout performs finite-cell selection on the GPU for repeated calculations with an evolving stellar field.

The nominal ray count $N_{\rm ray}$ specifies the corresponding full-field regular grid, with spacings $(\Delta\theta_x,\Delta\theta_y)$. The scout uses every $k$th fine-grid vertex, so each scout cell contains $k\times k$ fine cells. We trace its corners through the current lens equation and select the cell if the resulting quadrilateral intersects the source field. Testing the quadrilateral rather than only its vertices retains cells whose mapped footprint crosses the source even when every corner lies outside it. Figure~\ref{fig:scout_ipm_schematic} illustrates this selection and the subsequent IPM calculation.

We search cells intersecting the circular stellar aperture and one neighbouring layer outside it. Within this search region, we also retain one layer of neighbours around every selected cell to guard against contributions missed by the coarse scout. For the adopted $k=2$, each scout cell spans $2\times2$ fine cells, so each guard layer is two fine cells wide.

At the first epoch, we use the denser $k=1$ scout to estimate a normalization correction for the faster $k=2$ selection. We add this correction uniformly to subsequent magnification maps before integrating the source brightness. It accounts for an approximately uniform missing contribution, but does not recover unresolved spatial structure. The dense selection is then discarded, and subsequent epochs use the faster scout with temporal reuse.

A fixed rectangular shooting region can exclude cells that contribute to the source flux. The resulting error is primarily a missing-flux bias, but the omitted contribution can change as the stars move and therefore alter the light-curve shape. Our scout instead selects contributing cells from across the full circular stellar aperture, whose size is set by the large-deflection criterion in Section~\ref{subsec:microlens_population}.

\subsection{Temporal scout reuse}
\label{subsec:temporal_scout_reuse}

Nearby epochs often receive flux from much the same lens-plane region, even as stellar motion changes the magnification within it. We exploit this by reusing the selected cells between scout evaluations while recalculating their lens mapping at every epoch.

We repeat the scout calculation every ten frames. Between successive scout evaluations, we retain every cell selected at either endpoint to accommodate motion of the contributing region. The selection at the end of one interval also serves as the starting selection for the next. Only the cell selection is reused. The lens mapping, magnification, and Jacobian are recalculated at every frame using the current stellar positions. Several epochs are processed together on the GPU, each with its own stellar field.

For the benchmark's 25-day map cadence, the scout evaluations are separated by 250 days. A cell that contributes only between the endpoints can still be missed, so the refresh interval must remain short enough to capture changes in the contributing region. The end-to-end benchmarks in Section~\ref{subsec:qualitative_results} validate the light-curve accuracy and caustic-crossing recovery with the adopted ten-frame scout interval.

\subsection{Biquadratic inverse polygon mapping}
\label{subsec:biquadratic_ipm}

IPM distributes the mapped area of each finite lens-plane cell among the source-plane pixels it intersects, rather than depositing its entire contribution at a single ray position \citep{Mediavilla2006,Mediavilla2011}. This area-mapping principle is also used in recent FMM--IPM implementations \citep{JimenezVicente2022,Weisenbach2025Maps}. Our implementation reconstructs curvature in the lens mapping from a small grid of traced rays. We then subdivide this reconstructed mapping more finely without additional lens-equation evaluations and calculate the contributions to source pixels together on the GPU.

Classical IPM traces the corners of a lens-plane cell and represents its mapped area with straight triangles. This approximation can miss curvature near critical curves. Within each fine cell retained by the scout, we instead evaluate the lens equation on an $(r+1)\times(r+1)$ grid, including its four already available corners. Our adopted $r=2$ supplies a $3\times3$ grid of nine traced nodes. We describe positions within the cell by normalized coordinates $(\xi,\zeta)$, each running from zero to one. The nodes lie at $0$, $1/2$, and $1$ along each axis. To reconstruct the mapping between them, we fit a tensor-product quadratic Lagrange interpolant,
\begin{equation}
 \widetilde{\boldsymbol{\beta}}(\xi,\zeta)
 =
 \sum_{a=0}^{2}\sum_{b=0}^{2}
 L_a(\xi)L_b(\zeta)\,\boldsymbol{\beta}_{ab},
 \qquad 0\leq\xi,\zeta\leq1,
\label{eq:biquadratic_source_mapping}
\end{equation}
where $\boldsymbol{\beta}_{ab}$ are the traced source-plane coordinates and $L_a$ are the one-dimensional quadratic Lagrange basis functions for these three node positions. The interpolant reproduces all nine traced values. It reconstructs the lens mapping, not the magnification, which is calculated from the mapped areas.

We evaluate the reconstructed mapping on a finer $(v+1)\times(v+1)$ virtual grid, with $v=4$ in the adopted configuration. These virtual vertices require no additional lens-equation evaluations. Each virtual lens-plane subcell is split into two triangles, whose vertices are mapped through the interpolant. Each triangle's lens-plane area is then distributed among source pixels in proportion to their overlap with its mapped triangle. The triangle--pixel overlap areas are calculated analytically. The resulting straight-sided triangles approximate the curved mapping without the more complicated calculation of intersections between curved cell boundaries and source pixels.

The parameters $r$ and $v$ therefore control different parts of the approximation. Increasing $r$ samples the physical lens equation at more positions and raises the interpolant's order. For example, changing $r$ from 2 to 4 replaces the nine-node biquadratic reconstruction with a 25-node biquartic reconstruction. Increasing $v$ instead subdivides the same reconstructed mapping more finely. At $r=2$, changing $v$ from 2 to 4 retains the nine traced nodes but increases the virtual grid from $3\times3$ to $5\times5$. Our adopted configuration uses $r=2$ and $v=4$.

Fig.~\ref{fig:scout_ipm_schematic} illustrates this calculation. A fused Triton kernel performs the virtual interpolation, triangle--pixel clipping, and accumulation together, avoiding the large intermediate arrays required by the corresponding PyTorch calculation \citep{Tillet2019Triton}. This accumulation of mapped areas into source pixels is the rasterization step. The same integrator can be applied to the full stellar aperture, a fixed rectangular region, or the scouted cells. Comparing these variants isolates the effects of spatial selection while holding the integration method fixed. Section~\ref{subsec:map_convergence_rate} tests convergence with ray count, including the adopted $10^7$-ray, $r=2$, $v=4$ configuration, against the high-resolution $10^9$-ray, $r=2$, $v=8$ reference.

\subsection{From magnification maps to light curves}
\label{subsec:map_flux_batching}

IPM divides the image-plane area deposited in each source pixel by that pixel's area to obtain its magnification. We multiply the resulting map by the source brightness and sum over pixels to calculate the flux using equation~\eqref{eq:finite_source_flux}. Several epochs and source bands are processed together on the GPU, with each epoch retaining its own magnification map and time-dependent source brightness. We can also process several independent light curves together, each with its own stellar field and source history. Only the final products are transferred to the CPU.

\subsection{Critical curves, caustics, and binary caustic labels}
\label{subsec:anchor_gauge}

We identify when a caustic passes across the source centre and measure the distance to the nearest caustic. Rather than tracing the entire caustic network, we construct only the caustics that can affect the requested source field. We then count intersections with these caustics to assign a binary label, using several reference points to make the classification reliable and consistent between frames.

The critical-curve calculation evaluates $\det\mathbf{J}(\boldsymbol{\theta},t)$, with $\mathbf J$ defined in equation~\eqref{eq:local_lens_jacobian}, using analytic derivatives of the local-exact/far-field deflection. We first search a coarse grid for cells with a determinant sign change or unusually small $\lvert\det\mathbf J\rvert$. We add two neighbouring layers of coarse cells around these candidates, then evaluate the determinant at high resolution only in the selected regions. A complementary coarse scout searches the remainder of the stellar aperture for critical cells whose mapped footprints approach the source. Together, these searches retain relevant critical structure both near the nominal macroimage footprint and elsewhere in the stellar field.

The adopted determinant grid is $8192\times8192$, with a $512\times512$ discovery grid, corresponding to a factor of 16 along each axis. The benchmark reference uses $16384\times16384$. Rectangular controls retain the same physical pixel spacing as the $8192\times8192$ full-field grid, with their grid dimensions adjusted to the rectangular footprint.

A batched Triton marching-squares calculation locates the zero-determinant contours as image-plane line segments. Ambiguous saddle configurations are resolved using the bilinear asymptotic decider \citep{NielsonHamann1991}. We ray trace the segment endpoints to obtain the source-plane caustics $\mathcal C_\beta(t)$ defined in equation~\eqref{eq:critical_caustic}. Open or incomplete components at the computational boundary are flagged for the subsequent label calculation.

\begin{figure}
    \centering
    \includegraphics[width=\linewidth]{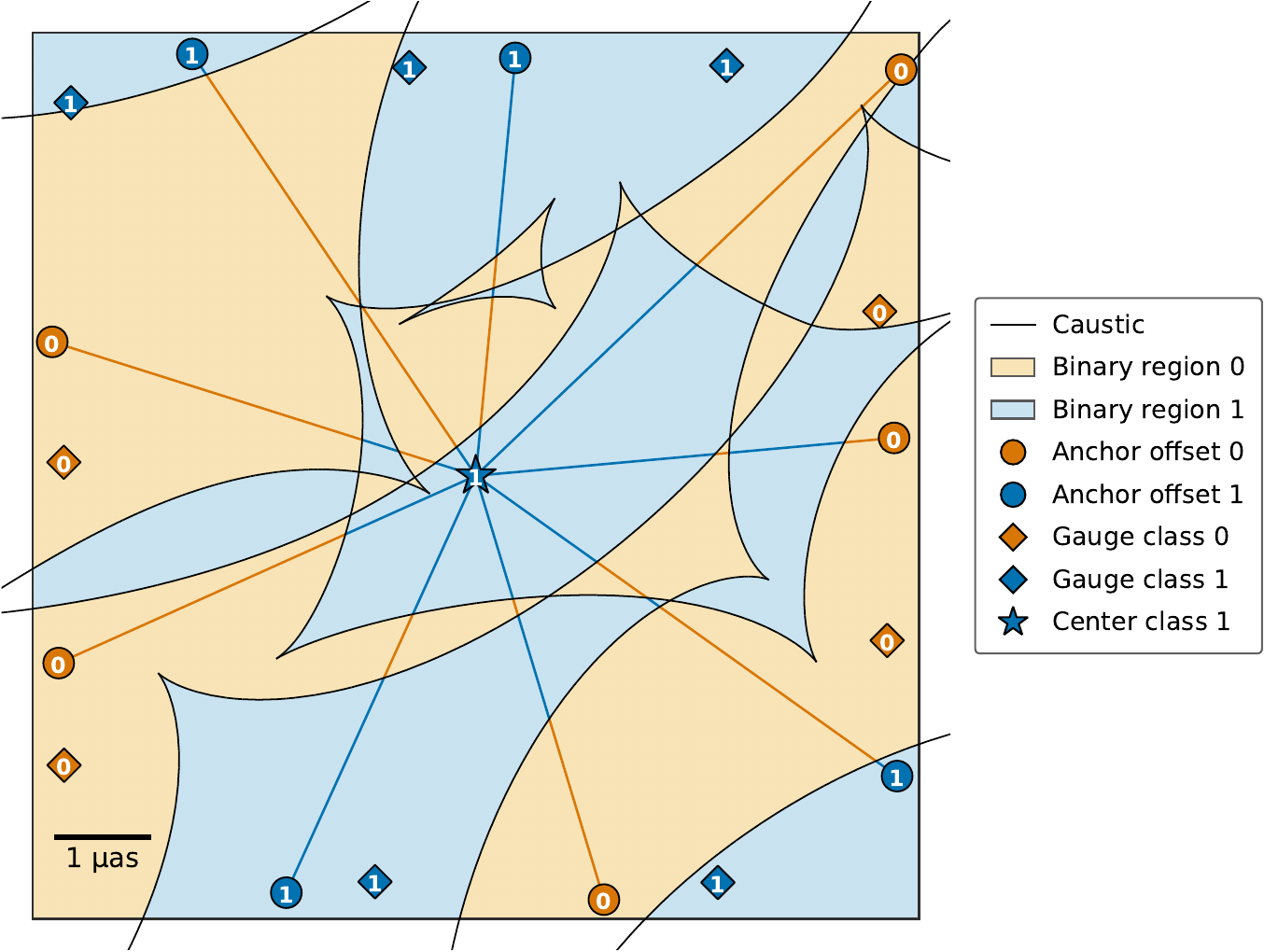}
    \caption{
Anchor--gauge binary caustic labels for one dynamical frame. Caustics divide the source plane into alternating binary regions. Anchors near the source-field boundary establish a common binary convention within each frame. Gauges keep that convention consistent between frames, with greater weight given to points farther from caustics. The two sets of reference points are spatially offset to avoid coincident paths and boundary degeneracies. Lines to the source centre illustrate the crossing count for one source position. The binary field uses the native $1024\times1024$ source grid.
    }
    \figurealttext{A source-plane caustic network separates alternating binary regions. Boundary anchors and interior gauges are connected to the source centre by probe paths, illustrating how crossing counts assign and align the binary caustic class.}
\label{fig:anchor_gauge}
\end{figure}

For a source-plane point away from a caustic, we define its binary class relative to a reference point, or \emph{anchor}, $\boldsymbol a$:
\begin{equation}
 \chi_{\rm c}(\boldsymbol{\beta},t)=
 N_{\rm cross}\!\left(
 \boldsymbol{a}\rightarrow\boldsymbol{\beta};
 \mathcal{C}_\beta(t)\right)\bmod 2,
\label{eq:binary_caustic_class}
\end{equation}
where $N_{\rm cross}$ counts transverse intersections between the straight reference segment and the caustics. Even and odd crossing counts give opposite labels, so adjacent caustic regions have different classes. Applying this calculation across the source plane produces the binary caustic map illustrated in Fig.~\ref{fig:anchor_gauge}. The labels zero and one have no absolute physical meaning, and their convention must be aligned between frames before changes can be interpreted as crossings.

A single reference segment can give an unreliable count if it passes close to a caustic vertex, is nearly tangent to a caustic, or encounters an incomplete or under-resolved segment. We therefore use nine anchors surrounding the source. Crossing counts between the anchors first place their classifications in a common binary convention. Each anchor then supplies a classification for the source position, and the majority determines its label. The different directions reduce sensitivity to a problem affecting one path.

We reject an anchor vote when its reference segment intersects a flagged incomplete component. The remaining anchors can still determine the class without requiring every caustic component to be fully contained in the computational field. Nine valid votes cannot tie. If rejected votes leave an exact tie, we assign class zero by a fixed convention and report the vote and validity counts so that the ambiguity remains identifiable.

To align the zero and one labels between successive frames, we use a second set of nine fixed reference points, called \emph{gauges}. Their votes select the convention that best preserves the preceding frame's labels. A gauge close to a caustic is unreliable because a small displacement of the discretized curve can change which side contains the point. We therefore prefer gauges farther from caustics in both frames and weight their votes by their distance to the nearest caustic. Using several gauges reduces the influence of any one point near a caustic or an incomplete segment. The anchors and gauges are placed near the perimeter of the source field, with small positional offsets to break symmetry, and remain fixed throughout the simulation, as illustrated in Fig.~\ref{fig:anchor_gauge}.

The binary class records only whether the reference path crosses an even or odd number of caustics. By comparison, a winding number counts how many times an oriented caustic curve wraps around a source position, including the direction of wrapping. The complete-field method of \citet{Weisenbach2025} traces the full critical-curve network and maps it to oriented caustics for this calculation. Our public code also supports complete winding maps, but the principal benchmarks use the less expensive source-centred binary labels.

Once the convention is aligned, we return the binary label, crossing history, and Euclidean distance to the nearest retained caustic segment. These quantities can be evaluated at any source-plane point, although the principal benchmark uses only the source centre. Two caustic crossings between successive epochs leave the binary label unchanged and would therefore be missed. This motivates a sufficiently short sampling interval. At the adopted 25-day cadence, our validation recovers all reference crossings (Section~\ref{subsec:qualitative_results}). A source-centre crossing is a geometric event marker, whereas the extended source's photometric response can begin earlier and depends on its full brightness profile.

\section{Validation and Performance Benchmarks}
\label{sec:validation_results}

\subsection{Benchmark sample, methods, and timing procedure}
\label{subsec:benchmark_sample}

\begin{table*}
\centering
\caption{Benchmark lens and field parameters\label{tab:benchmark_systems}}
\footnotesize
\begin{tabular*}{0.96\textwidth}{@{\extracolsep{\fill}}llcccccccc}
\hline
    System &
    Image &
    $(z_l,z_s)$ &
    $\kappa$ &
    $\gamma$ &
    $\phi_\gamma$ [deg] &
    $s$ &
    $N_\star$ &
    $L_{\rm lens}$ [$\mu\mathrm{as}$ ($\theta_{\rm E}$)] &
    $L_{\rm source}$ [$\mu\mathrm{as}$ ($\theta_{\rm E}$)] \\
\hline
 Q2237+0305 & A & $(0.0395,1.70)$ & 0.396 & 0.396 & 175 & 0.00 & 4263 & 793 (208) & 9.38 (2.46) \\
              & B &                   & 0.391 & 0.391 & 142 & 0.00 & 3805 & 754 (197) &  \\
              & C &                   & 0.715 & 0.715 & 69.1 & 0.00 & 3549 & 539 (141) &  \\
              & D &                   & 0.604 & 0.604 & 62.5 & 0.00 & 9568 & 962 (252) &  \\
 RX J1131$-$1231 & A & $(0.295,0.658)$ & 0.526 & 0.410 & 73.2 & 0.571 & 16840 & 638 (546) & 4.60 (3.94) \\
              & B &                   & 0.459 & 0.412 & 106 & 0.566 & 3352 & 303 (259) &  \\
              & C &                   & 0.487 & 0.306 & 39.5 & 0.586 & 1394 & 194 (166) &  \\
              & D &                   & 0.894 & 0.807 & 76.0 & 0.419 & 1056 & 105 (90.2) &  \\
 HE 0435$-$1223 & A & $(0.455,1.69)$ & 0.473 & 0.358 & 63.8 & 0.653 & 1214 & 195 (172) & 2.89 (2.54) \\
              & B &                   & 0.630 & 0.540 & 165 & 0.639 & 2208 & 224 (197) &  \\
              & C &                   & 0.494 & 0.327 & 88.7 & 0.666 & 1104 & 186 (164) &  \\
              & D &                   & 0.686 & 0.575 & 168 & 0.620 & 1274 & 159 (140) &  \\
\hline
\end{tabular*}
\par\vspace{4pt}
\begin{minipage}{0.98\textwidth}
\footnotesize\textsc{Note}: Blank redshift and source-field entries repeat the values from the first macroimage of each system. Here $s$ is the smooth-matter fraction and $\phi_\gamma$ is the local shear position angle, modulo $180^\circ$ and measured counterclockwise from the positive horizontal coordinate. We use the tangential approximation derived from the published macroimage and lens-centre astrometry, rather than treating $\phi_\gamma$ as an independently measured model parameter. The Q2237 macro parameters are from \citet{Tuntsov2024Planets}. RX~J1131 and HE~0435 use \citet{Chen2019SHARP}, with $s=1-\kappa_\star/\kappa$. Their astrometry is from \citet{Wertz2014Q2237}, \citet{WittMao2015}, and \citet{FadelyKeeton2012}, respectively. The system-level source tuples $(\log_{10}(M_{\rm BH}/M_\odot),a,i,\lambda_{\rm Edd},\epsilon)$ are $(9.08,0.74,10^\circ,0.34,20)$ for Q2237, $(8.30,0.87,17^\circ,0.07,10)$ for RX~J1131, and $(8.36,0.70,30^\circ,0.19,20)$ for HE~0435. For Q2237, the black-hole mass and spin follow \citet{Reynolds2014EinsteinCross}, $i=10^\circ$ is a fiducial consistent with their upper limit, $\lambda_{\rm Edd}=0.34$ is a fiducial motivated by \citet{Berdina2021}, who assumed a lower black-hole mass, and $\epsilon=20$ is a typical value. For RX~J1131, the mass, spin, and inclination follow \citet{Reis2014}, while $\lambda_{\rm Edd}=0.07$ is an adopted fiducial and $\epsilon=10$ is a typical value. For HE~0435, the mass follows \citet{Melo2023}, while $a=0.70$, $i=30^\circ$, $\lambda_{\rm Edd}=0.19$ are adopted fiducials, and $\epsilon=20$ is a typical value. We set the shared fiducial source-plane disc position angle to $\phi_{\rm disc}=0^\circ$. This is not an observational constraint or the shear direction. Also, $H_{\rm lamp}/R_g=\epsilon+R_{\rm ISCO}/R_g$. Shared source settings are $\beta=0.75$, $f_{\rm col}=1$, and $f_{\rm lamp}=0.10$. The Einstein-radius conversion adopts the benchmark mean microlens mass $\langle M_\star\rangle=0.3\,M_\odot$ with $M_{\max}/M_{\min}=100$ and Salpeter slope $\alpha=2.35$. Every macroimage uses 147 frames at 25-day intervals.
Here $f_{\rm lamp}$ denotes the albedo-weighted coronal X-ray luminosity in Eddington units.
\end{minipage}
\end{table*}

For each macroimage, we compare methods using the same physical configuration and stellar realization on the same RTX 5070 Ti GPU. Table~\ref{tab:benchmark_systems} gives the lens, source, and field parameters. The 12 macroimages include both parities and contain 1,056--16,840 microlenses. They span a broad range in local convergence, shear, smooth-matter fraction, and number of cells contributing to the source. These systems provide a compact set of contrasting stellar fields with published lens and accretion-disc constraints. We adopt reasonable fiducial values where constraints are weak, since our purpose is to demonstrate numerical accuracy and runtime rather than model these systems individually.

Benchmark values are medians across the 12 macroimages, with superscript and subscript offsets reaching the observed 84th and 16th percentiles, respectively. These offsets describe case-to-case spread, not confidence intervals on the median. The four images within each lens system are related, and the benchmark is not a population sample.

\label{subsec:benchmark_contract}

Throughout the comparisons, $N_{\rm ray}$ denotes the number of launched rays for IRS and the nominal number of image-plane grid samples for IPM. Our adopted source-scouted biquadratic IPM configuration uses $N_{\rm ray}=10^7$, $k=2$, $r=2$, and $v=4$. It reaches approximately the map accuracy of the calculation of \citet{Weisenbach2025Maps} at its fiducial setting of one ray per output source pixel along the $y$ direction and a 1 per cent light-loss tolerance. This setting is not directly equivalent to our nominal ray count, so we compare the measured accuracy rather than assign an equivalent ray budget.

Our internal IRS controls launch rays on a regular Cartesian lattice and use the same lens fields, deflection solver, nominal ray counts, and batching settings as the corresponding IPM controls. The internal rectangular controls and GPU-D share a shear-aligned 1 per cent light-loss window expanded by 10 per cent along each axis, while retaining the complete circular microlens field. GPU-D uses pseudorandom IRS with the same nominal $10^7$-ray budget. The calculation of \citet{Weisenbach2025Maps} uses its native shooting window at the same nominal light-loss tolerance.

We report root-mean-square errors (RMSEs) for light curves and maps, with fractional map RMSE defined in equation~\eqref{eq:map_fractional_rmse}. The dynamical map and light-curve benchmarks use a reference with $10^8$ nominal rays per epoch, $r=2$, and $v=8$ at all 147 epochs. The snapshot maps in Fig.~\ref{fig:method_comparison} and convergence tests in Fig.~\ref{fig:map_convergence_rate} use a $10^9$-ray reference with the same refinement settings. The internal candidate and reference maps share the same $1024^2$ source grid and physical pixel boundaries.

Timings exclude compilation and autotuning. The light-curve time $t_{\rm LC}$ includes magnification-map calculation and integration of the source brightness over all 147 epochs. The time $t_{\rm LC+labels}$ also includes determinant evaluation, caustic construction, and source-centre labels. Internal IPM calculations process cells in groups of 524,288.

\citet{Weisenbach2025Maps} calculates one map at a time. We measure their light-curve runtime and accuracy over all epochs, and compare their maps with references evaluated on the same complete, shear-aligned native grid. For this method and the direct-deflection controls, we measure the caustic-stage cost at the first epoch and scale it to the full sequence. Our labelled calculation returns source-centre labels, whereas the timing for \citet{Weisenbach2025} includes construction of the complete critical-curve and caustic network.

GPU-D timings include initialization, memory allocation, transfer of the microlens data to the GPU, ray shooting, and accumulation into source-plane pixels. GPU-D does not calculate caustics or labels.

\subsection{Accuracy, runtime, and example map residuals}
\label{subsec:qualitative_results}

\begin{figure*}
    \centering
    \includegraphics[width=\linewidth]{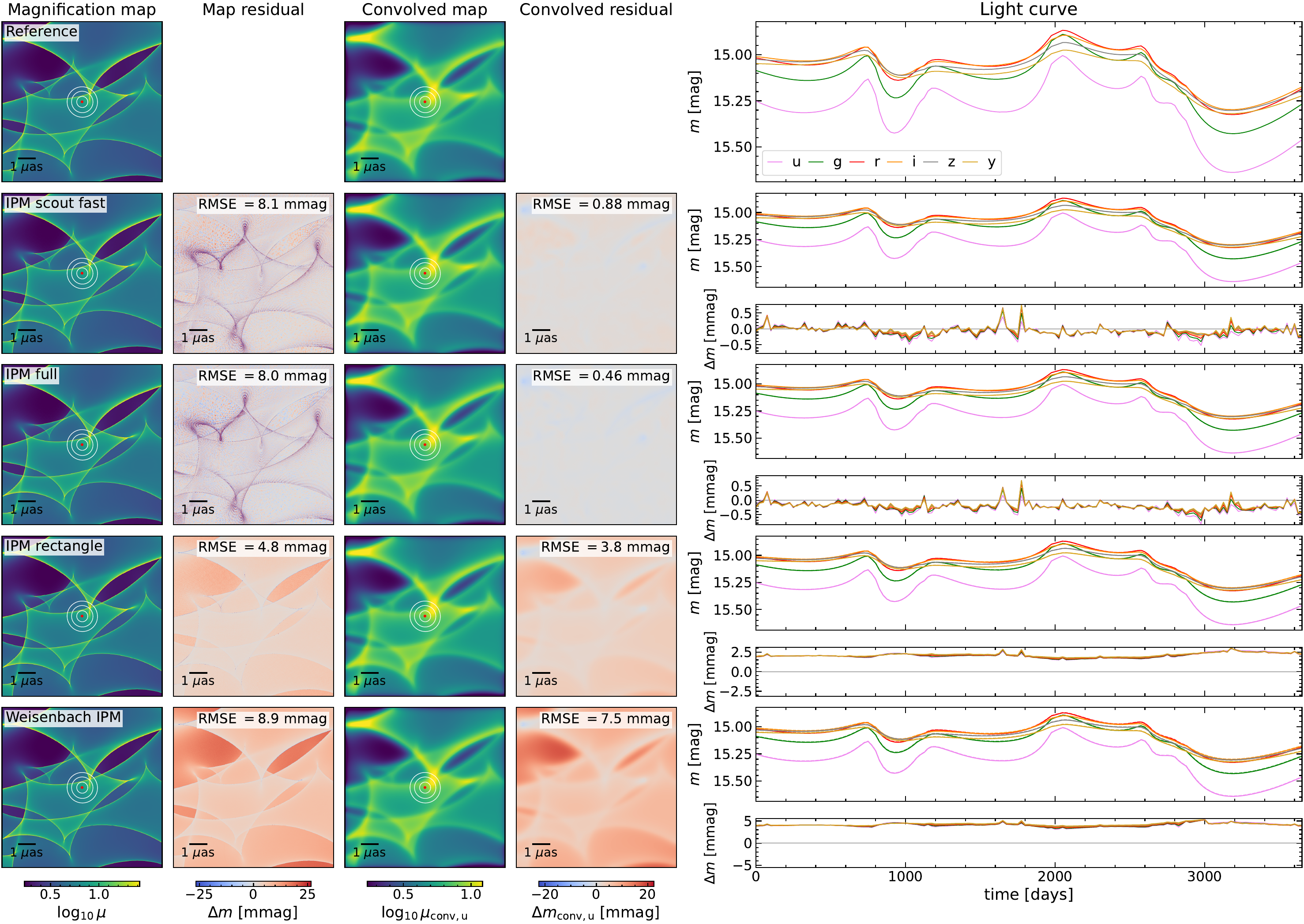}
    \caption{
Comparison of the high-quality reference, the fast source-scouted IPM calculation, full-field and rectangular IPM controls, and Weisenbach IPM for Q2237+0305 image B. Columns show the magnification map and signed residual, $u$-band source-convolved map and residual, and six-band light curve with its residuals below. Map residuals use $\Delta m=-2.5\log_{10}(\mu/\mu_{\rm ref})$ on signed linear millimagnitude scales shared within each column. All maps are evaluated on the same shear-aligned source-plane square. The map reference uses $10^9$ rays, while the full light-curve reference uses $10^8$ rays per epoch, both with $r=2$ and $v=8$. All internal IPM calculations use $N_{\rm ray}=10^7$, $r=2$, and $v=4$, with $k=2$ for the source-scouted calculation.
    }
    \figurealttext{Rows compare the reference and four microlensing methods. Columns show raw and source-convolved magnification maps, their residuals, and six-band light curves. The maps look similar, while residual panels expose method-dependent offsets and small-scale differences.}
\label{fig:method_comparison}
\end{figure*}

Fig.~\ref{fig:method_comparison} shows magnification maps, finite-source light curves, and residuals for the principal IPM methods before we summarize the full 12-macroimage comparison. The source-scouted and full-field light curves closely follow the reference in all six bands, with residuals on the sub-millimagnitude scale in this example.

\begin{figure*}
    \centering
    \includegraphics[width=\linewidth]{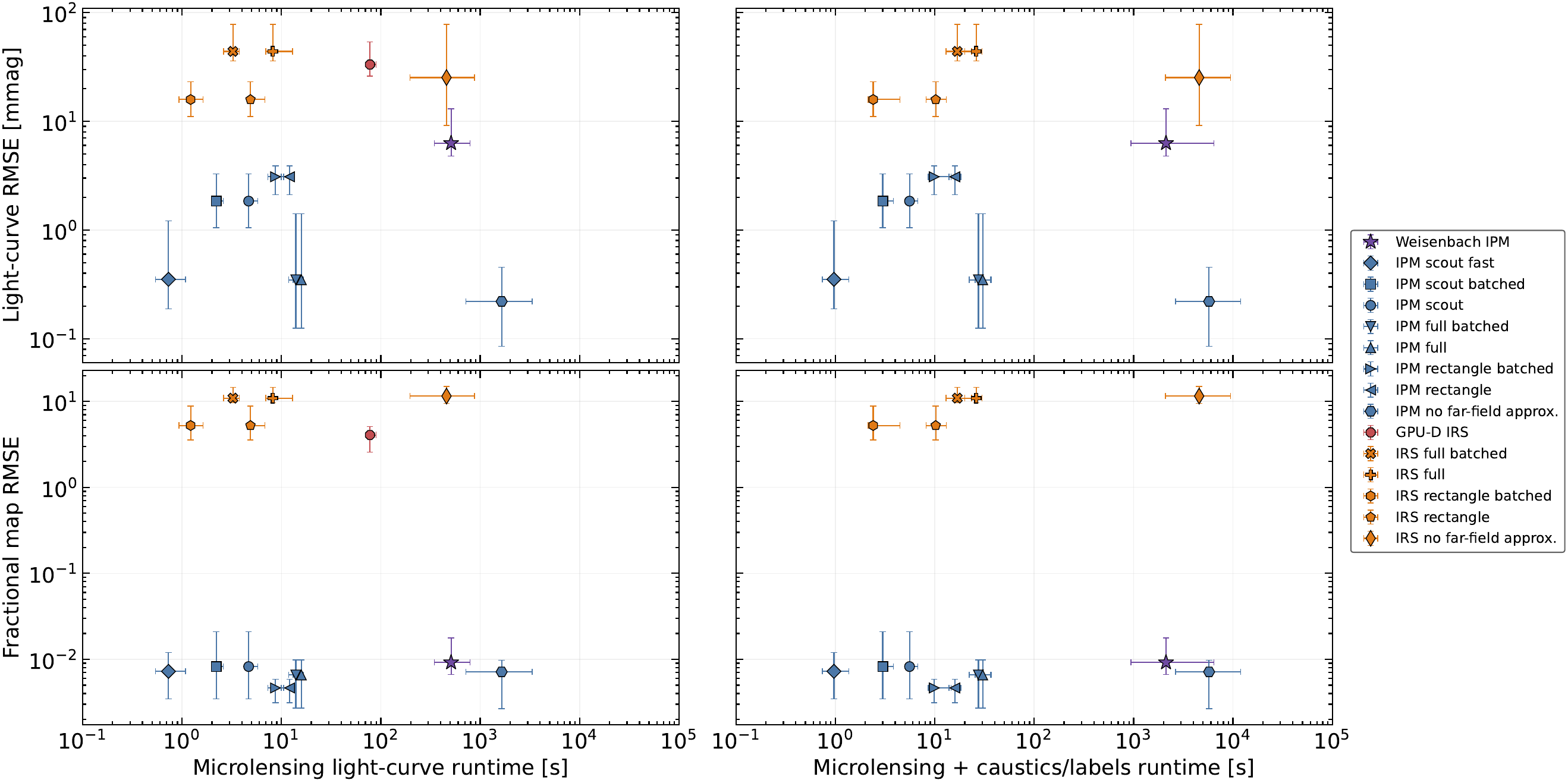}
    \caption{
Accuracy and runtime across 3 systems and 12 macroimages for complete 147-frame light curves and magnification maps. The left panels use light-curve-only runtime and the right panels include caustics and source-centre labels, following the timing definitions in Section~\ref{subsec:benchmark_contract}. The reference uses $10^8$ nominal rays per epoch, $r=2$, and $v=8$. GPU-D appears only in the left panels because it provides no labels. Full-field and rectangular controls use the same nominal $10^7$ samples. The rectangle samples its smaller footprint more densely and can have lower fractional map RMSE, while omitted contributions outside it increase light-curve RMSE. Rectangular determinant grids retain the full-field physical pixel spacing without oversampling.
    }
    \figurealttext{Four log-scale panels compare light-curve and map errors against runtime, without and with caustic labels. Points and error bars summarise the macroimage sample. Source-scouted IPM combines low errors with shorter runtimes than the full-field and external comparisons.}
\label{fig:accuracy_runtime}
\end{figure*}

In Fig.~\ref{fig:accuracy_runtime} and Table~\ref{tab:accuracy_runtime}, ``full'' and ``rectangle'' denote sampling over the full lens-plane field and the restricted rectangular shooting region, respectively. IRS deposits individual rays, whereas IPM distributes mapped cell areas among source pixels. ``IPM scout'' selects source-reaching cells independently at each epoch. ``IPM scout batched'' evaluates these independent frames in temporal batches without reusing selected cells or sharing normalization corrections. ``IPM scout fast'' additionally reuses selected cells between scout refreshes and applies a fixed normalization correction measured at frame zero.

\begin{table*}
\centering
\caption{Light-curve and magnification-map benchmark\label{tab:accuracy_runtime}}
\footnotesize
\setlength{\tabcolsep}{4pt}
\begin{tabular*}{\textwidth}{@{\extracolsep{\fill}}lcccccc@{}}
\hline
    Method &
    Batch (LC/labels) &
    $t_{\rm LC}$ [s] &
    $t_{\rm LC+labels}$ [s] &
    LC RMSE [mmag] &
    Frac. map RMSE &
    Map RMSE [mmag] \\
\hline
Weisenbach IPM & 1 & $510_{-160}^{+280}$ & $2100_{-1200}^{+4300}$ & $6.3_{-1.5}^{+6.7}$ & $9.2_{-2.5}^{+8.6}\times10^{-3}$ & $10_{-2.7}^{+9.4}$ \\
IPM scout fast & 49/30 & $0.73_{-0.19}^{+0.35}$ & $0.97_{-0.23}^{+0.40}$ & $0.35_{-0.16}^{+0.87}$ & $7.2_{-3.8}^{+4.7}\times10^{-3}$ & $8.0_{-4.2}^{+4.0}$ \\
IPM scout fast (multi-curve) & 49/30 & $0.68_{-0.20}^{+0.30}$ & $0.88_{-0.20}^{+0.34}$ & $0.35_{-0.16}^{+0.87}$ & $7.2_{-3.8}^{+4.7}\times10^{-3}$ & $8.0_{-4.2}^{+4.0}$ \\
IPM scout batched & 49 & $2.2_{-0.17}^{+0.39}$ & $3.0_{-0.32}^{+0.83}$ & $1.8_{-0.79}^{+1.4}$ & $8.2_{-4.8}^{+13}\times10^{-3}$ & $8.1_{-4.2}^{+4.0}$ \\
IPM scout & 1 & $4.7_{-0.030}^{+1.1}$ & $5.6_{-0.24}^{+1.2}$ & $1.8_{-0.79}^{+1.4}$ & $8.2_{-4.8}^{+13}\times10^{-3}$ & $8.1_{-4.2}^{+4.0}$ \\
IPM full batched & 10 & $14_{-2.1}^{+2.8}$ & $28_{-5.4}^{+9.0}$ & $0.35_{-0.22}^{+1.1}$ & $6.6_{-3.9}^{+3.2}\times10^{-3}$ & $7.4_{-3.9}^{+3.1}$ \\
IPM full & 1 & $16_{-2.7}^{+0.66}$ & $30_{-4.8}^{+6.5}$ & $0.35_{-0.22}^{+1.1}$ & $6.6_{-3.9}^{+3.2}\times10^{-3}$ & $7.4_{-3.9}^{+3.1}$ \\
IPM rectangle batched & 49/30 & $8.7_{-1.4}^{+1.8}$ & $9.9_{-1.3}^{+4.0}$ & $3.1_{-0.98}^{+0.79}$ & $4.6_{-1.5}^{+1.2}\times10^{-3}$ & $5.4_{-1.6}^{+1.2}$ \\
IPM rectangle & 1 & $12_{-2.1}^{+1.5}$ & $16_{-1.3}^{+2.4}$ & $3.1_{-0.98}^{+0.79}$ & $4.6_{-1.5}^{+1.2}\times10^{-3}$ & $5.4_{-1.6}^{+1.2}$ \\
IPM no far-field approx. & 1 & $1600_{-920}^{+1700}$ & $5800_{-3100}^{+6200}$ & $0.22_{-0.14}^{+0.23}$ & $7.1_{-4.5}^{+2.6}\times10^{-3}$ & $8.1_{-4.7}^{+2.4}$ \\
GPU-D IRS & 1 & $78_{-1.2}^{+12}$ & --- & $33_{-7.2}^{+20}$ & $4.1_{-1.5}^{+1.0}\times10^{0}$ & $2500_{-680}^{+800}$ \\
IRS full batched & 49/10 & $3.3_{-0.63}^{+0.54}$ & $17_{-3.9}^{+6.6}$ & $44_{-8.0}^{+34}$ & $1.1_{-0.13}^{+0.36}\times10^{1}$ & $4600_{-330}^{+720}$ \\
IRS full & 1 & $8.2_{-1.2}^{+4.7}$ & $26_{-6.0}^{+3.8}$ & $44_{-8.0}^{+34}$ & $1.1_{-0.13}^{+0.36}\times10^{1}$ & $4600_{-330}^{+720}$ \\
IRS rectangle batched & 49/30 & $1.2_{-0.29}^{+0.40}$ & $2.4_{-0.26}^{+2.1}$ & $16_{-4.8}^{+7.3}$ & $5.2_{-1.7}^{+3.6}\times10^{0}$ & $2900_{-830}^{+490}$ \\
IRS rectangle & 1 & $4.9_{-0.34}^{+1.9}$ & $10_{-2.1}^{+2.9}$ & $16_{-4.8}^{+7.3}$ & $5.2_{-1.7}^{+3.6}\times10^{0}$ & $2900_{-830}^{+490}$ \\
IRS no far-field approx. & 1 & $460_{-260}^{+420}$ & $4600_{-2500}^{+4900}$ & $25_{-16}^{+53}$ & $1.2_{-0.21}^{+0.34}\times10^{1}$ & $4600_{-320}^{+710}$ \\
\hline
\end{tabular*}
\par\vspace{4pt}
\begin{minipage}{\textwidth}
\footnotesize\textsc{Note}: 
 The reference and timing definitions are given in Section~\ref{subsec:benchmark_contract}. Batch is the number of temporal frames evaluated jointly for light curves alone and with labels. A single value applies to both products. The fast path uses batches 49 and 30, respectively, because their throughput optima differ. For the multi-curve row, runtime is the total elapsed time divided by the number of curves, not the time required to calculate one curve alone. This row is omitted from Fig.~\ref{fig:accuracy_runtime}. Light-curve RMSE covers the complete light curve. Fractional map RMSE uses pixelwise fractional magnification residuals. Magnitude-map RMSE uses $-2.5\log_{10}(\mu/\mu_{\rm ref})$ only where both maps have positive magnification, excluding empty IRS pixels.
\end{minipage}
\end{table*}

At the adopted ray count, IRS sparsely samples the small source field, giving much larger raw-pixel errors than finite-source light-curve errors. The independent GPU-D ray-count test below checks their decrease with increasing ray count against the expected counting-noise scaling.

Fig.~\ref{fig:accuracy_runtime} and Table~\ref{tab:accuracy_runtime} summarize the accuracy--runtime trade-off across all 12 macroimages. The source-scouted method approaches the full-field IPM accuracy at a fraction of its cost, both for light curves alone and when caustics and labels are included. For the adopted fast simulator configuration, the median 147-frame light-curve runtime is $0.73^{+0.35}_{-0.19}\,$s. Including determinant evaluation, caustic construction, and source-centre labels increases the median runtime to $0.97^{+0.40}_{-0.23}\,$s. Relative to the all-frame reference, the corresponding median light-curve RMSE is $0.35^{+0.87}_{-0.16}\,$mmag and the raw fractional map RMSE is $7.2^{+4.7}_{-3.8}\times10^{-3}$. Across all 12 macroimages, the adopted labels recover all 41 reference source-centred caustic crossings, and every gauge signature agrees with the reference. The batched full-field IPM control has comparable median light-curve accuracy ($0.35\,$mmag) and slightly lower fractional map RMSE ($6.6\times10^{-3}$) but requires $14\,$s for the light curve, illustrating that cell selection and temporal reuse, rather than a different flux observable, drive the practical speedup.

Large simulation sets can also batch independent stellar realizations. To isolate the scheduling gain, we ran up to three separate calculations concurrently with the same seeded inputs as the single-curve control, without sharing maps or fluxes. The memory-heaviest RX~J1131$-$1231 images use one or two curves. The amortized median runtimes are $0.68^{+0.30}_{-0.20}\,$s per curve without labels and $0.88^{+0.34}_{-0.20}\,$s with labels, giving median throughput gains of $1.11^{+0.014}_{-0.041}$ and $1.12^{+0.064}_{-0.052}$ times, respectively. These are throughput measurements rather than single-curve latencies. Larger GPUs can evaluate more independent curves concurrently and should increase this advantage.

The calculation of \citet{Weisenbach2025Maps} takes a median $510^{+280}_{-160}\,$s for the complete light curve, with a median light-curve RMSE of $6.3^{+6.7}_{-1.5}\,$mmag and fractional map RMSE of $9.2^{+8.6}_{-2.5}\times10^{-3}$. Including the extrapolated cost of caustic construction \citep{Weisenbach2025}, we expect the complete calculation to take $2.1^{+4.3}_{-1.2}\times10^3\,$s.

With labels included and independent light curves processed together, our method achieves a median throughput advantage of $2.5^{+2.8}_{-1.2}\times10^3$ over this combined calculation. We calculate the runtime ratio for each macroimage before taking the median, rather than dividing the median runtimes.

For comparison with direct pseudorandom-ray shooting, we also ran the published GPU-D code \citep{Thompson2010,Bate2010,VernardosFluke2014Cloud}. Its rays were launched from the adopted 1 per cent light-loss rectangle while retaining the complete circular microlens field. At $10^7$ rays, the median complete process time across the 12 macroimages was $78^{+12}_{-1.2}\,$s for 147 independently calculated maps. We then measured frame zero for every macroimage at $N_{\rm ray}=\{10^7,3\times10^7,10^8,3\times10^8,10^9\}$. Free fits to the two lowest ray counts give a median slope of $-0.496^{+0.001}_{-0.002}$. We therefore adopt the theoretical counting-noise relation separately for every macroimage,
\begin{equation}
    \epsilon_{{\rm GPU-D},i}=A_i
    \left(\frac{N_{\rm ray}}{10^9}\right)^{-1/2},
    \label{eq:gpud_map_convergence}
\end{equation}
where $A_i$ is normalized with the $10^7$ and $3\times10^7$ ray measurements for macroimage $i$. The complete GPU-D process time per map is well described over the $10^7$--$3\times10^8$ interval by
\begin{equation}
    t_{{\rm GPU-D},i}=a_i+b_iN_{\rm ray},
    \label{eq:gpud_runtime_scaling}
\end{equation}
where $a_i$ and $b_i$ are the fitted process-time intercept and cost per ray for macroimage $i$. The median measured coefficient is $b_i=1.7\times10^{-8}\,$s per ray. Matching each macroimage's fast-IPM raw fractional map RMSE requires a median $3.5^{+4.0}_{-2.4}\times10^{12}$ rays. Equations \eqref{eq:gpud_map_convergence} and \eqref{eq:gpud_runtime_scaling} predict a median $100^{+100}_{-63}$ days for 147 maps. Pairing each extrapolated time with the corresponding fast-IPM throughput gives a median speed ratio of $1.4^{+2.2}_{-1.1}\times10^7$.

The measured errors follow the $N_{\rm ray}^{-1/2}$ relation through $10^9$ rays, where all 12 cases remain within 0.7 per cent of the prediction normalized at the two lowest ray counts. No finite-support floor is resolved over this range. The matched-accuracy estimate extrapolates this counting-noise relation far beyond the measured ray counts and assumes that missing rays outside the fixed shooting window do not limit the accuracy. It is therefore an idealized comparison favourable to GPU-D, not a demonstration that the fixed window reaches the fast-IPM accuracy. For context, the median extrapolated runtime is $16^{+17}_{-10}$ GPU-hours per map. 

Across independent timing repeats, individual median light-curve runtimes range from $0.50\,$s for HE~0435$-$1223 A to $1.8\,$s for RX~J1131$-$1231 A. Including caustics and labels, they range from $0.73\,$s to $2.1\,$s for the same two macroimages. Runtimes vary with the lens parameters, number of stars, aperture size, and number and distribution of cells contributing to the source. The sample is too small to infer a separate scaling with parity or stellar density.

Synchronized timings of individual stages in the same benchmark calculations across the 12 macroimages show where the calculation spends its time. For the light-curve-only workload, the median fractions are $13\,\mathrm{per\,cent}$ for the local-exact/far-field calculation, $15\,\mathrm{per\,cent}$ for source scouting and correction, $68\,\mathrm{per\,cent}$ for shared ray tracing, interpolation, and IPM rasterization, and $2.2\,\mathrm{per\,cent}$ for source evaluation and flux integration. Scheduling accounts for less than $1\,\mathrm{per\,cent}$. For the separately optimized labelled workload, which uses a total temporal batch of 30, the corresponding medians are $12\,\mathrm{per\,cent}$, $11\,\mathrm{per\,cent}$, $51\,\mathrm{per\,cent}$, and $1.8\,\mathrm{per\,cent}$. Caustic construction contributes $16\,\mathrm{per\,cent}$, while source-centre labelling and other computational overhead contribute $7.2\,\mathrm{per\,cent}$. These are medians of per-macroimage fractions and need not sum exactly to 100 per cent.

\subsection{Ray-count convergence of the magnification map}
\label{subsec:map_convergence_rate}

We measure the pixelwise fractional map RMSE as
\begin{equation}
 \epsilon_{\mu}=\left\langle
 \left(\frac{\mu_N-\mu_{\rm ref}}{\mu_{\rm ref}}\right)^2
 \right\rangle^{1/2},
\label{eq:map_fractional_rmse}
\end{equation}
where the average is over the compared source-plane pixels. At fixed map geometry, $\mu_N$ is the map calculated with $N_{\rm ray}$ rays and $\mu_{\rm ref}$ is the reference map on the same source-plane grid. This statistic tests relative magnification accuracy across both demagnified and magnified regions. Because each residual is divided by its reference-pixel value, it gives low-magnification pixels greater leverage than a globally normalized $L_2$ error. We write the ray-count dependence as $\epsilon_{\mu}\propto N_{\rm ray}^{p}$.

Independent random-ray deposition has the familiar Poisson scaling $p=-1/2$. Regular-lattice IRS can approach $p=-3/4$ when source-pixel boundary cells dominate \citep{Kayser1986,Mediavilla2011}, with a measured relative-error slope of $-0.79\pm0.01$ in \citet{Mediavilla2006}. \citet{Zheng2022GPU} reported an intermediate slope near $-2/3$. These exponents are expressed in rays per source pixel, which is proportional to the total ray count for the fixed field of view and output grid used below.

IRS assigns the full weight of an image-plane cell to its centre ray, whereas IPM transforms the cell boundary and apportions its mapped area among all intersected source pixels. The practical piecewise-affine IPM error is therefore a deterministic finite-cell or linearization error whose scaling depends on the magnification gradients, cell-to-pixel size, critical-curve curvature, and critical-cell treatment \citep{Mediavilla2011}. Our biquadratic reconstruction is integrated using straight-sided subtriangles.

A Taylor argument for affine IPM gives an idealized $N_{\rm ray}^{-1}$ convergence rate in smooth regions. Let $h$ denote the image-plane cell spacing. At fixed field of view, the number of traced lattice vertices scales as $N_{\rm ray}\propto h^{-2}$. Away from point-mass poles, the lens mapping can be expanded locally as
\begin{equation}
 \boldsymbol{\beta}(\boldsymbol{\theta}+\delta\boldsymbol{\theta}) =
 \boldsymbol{\beta}(\boldsymbol{\theta})
 + \mathbf{J}(\boldsymbol{\theta})\,\delta\boldsymbol{\theta}
 + \frac{1}{2}\mathbf{H}(\boldsymbol{\theta})
   [\delta\boldsymbol{\theta},\delta\boldsymbol{\theta}]
 + \mathcal{O}(h^3).
 \label{eq:ipm_local_taylor}
\end{equation}
Here $\mathbf{J}$ and $\mathbf{H}$ are the Jacobian and Hessian of the two-dimensional lens mapping. An affine mapped triangle retains the constant and linear terms. The Hessian term gives the leading $\mathcal{O}(h^2)$ displacement of the true mapped boundary. This displacement produces an $\mathcal{O}(h^3)$ area error per boundary cell. Summing over the $\mathcal{O}(h^{-1})$ cells intersected by a smooth source-pixel boundary gives the idealized integrated error
\begin{equation}
 \epsilon_{\rm affine\ IPM}=\mathcal{O}(h^2)
 =\mathcal{O}(N_{\rm ray}^{-1}).
 \label{eq:ipm_nominal_convergence}
\end{equation}
The exponent $-1$ is an idealized smooth-region rate, not a universal IPM upper bound. Approximate polygon filling, unresolved critical cells, and finite-field selection can make the observed rate shallower. The isolated point-lens experiment in Appendix~\ref{app:point_lens_convergence} checks the expected Monte Carlo IRS and IPM rates against the analytic solution, without finite-reference uncertainty. IPM also has a lower error normalization. \citet{Mediavilla2006} found that approximately one mapped cell per unlensed pixel was about an order of magnitude more accurate than IRS with roughly 1024 rays per unlensed pixel. This advantage decreases with source convolution \citep{Mediavilla2011}, making area mapping most valuable for raw maps and compact sources.

The nominal $N_{\rm ray}$ specifies the full-field lattice before scouting, not the number of lens-equation evaluations actually performed. The convergence experiment fixes the stellar field, source-plane field of view, $1024^2$ output grid, scout ratio $k=2$, true candidate refinement $r=2$, and biquadratic virtual raster refinement $v=4$, and varies only $N_{\rm ray}=\{1,2.5,5,10,20,50,100\}$ million. Fixed $r$ and $v$ affect the work and error normalization, not the idealized affine $h^2$ rate. Changing the scout ratio $k$ changes the selected cells and need not preserve the measured slope. Every candidate and the independent $10^9$-ray reference use the same conservative far-field settings: a $32\times32$ nodal lattice, $\chi_{\rm near}=1.5$, and an order-ten expansion about each cell centre. The reference uses $k=2$, $r=2$, and the finer virtual raster refinement $v=8$. Its higher ray count reduces finite-reference error when testing candidates up to $10^8$ rays.

\begin{figure*}
    \centering
    \includegraphics[width=\linewidth]{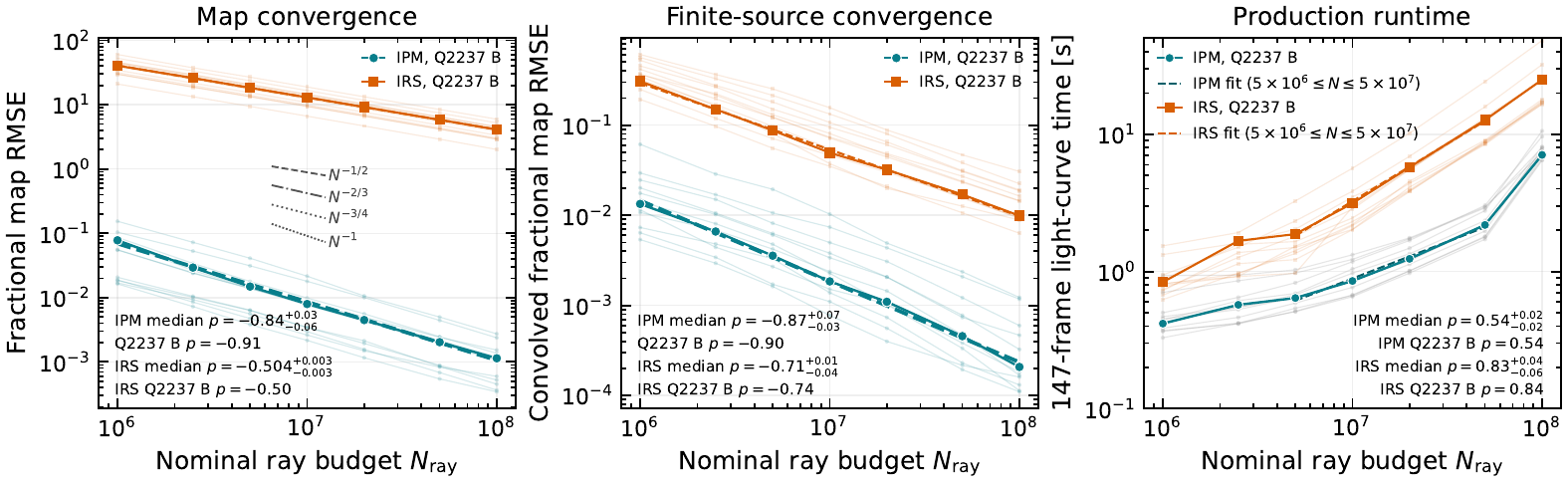}
    \caption{
Ray-count convergence and computational cost. Candidate selective biquadratic IPM calculations hold $k=2$, $r=2$, and $v=4$ fixed and use $N_{\rm ray}=\{1,2.5,5,10,20,50,100\}$ million. The left panels show raw and source-convolved fractional map RMSE against the matched $10^{9}$-ray, true-$r=2$, $v=8$ reference, with Q2237+0305 image B highlighted. Both IPM and matched IRS using the same local-exact/far-field deflection solver are evaluated for all 12 macroimages. The other 11 trajectories are shown faintly in the corresponding method colour. The right panel reports the runtime measured after compilation and warm-up for a 147-frame light curve using fixed local-exact/far-field solver settings. The temporal batch size is optimized for each method, macroimage, and ray count, with one timing per point. The runtime power law is fit empirically over $5\times10^6\leq N_{\rm ray}\leq5\times10^7$ near the $10^7$-ray operating point. The $10^6$ and $2.5\times10^6$ points are displayed but excluded because fixed launch overhead matters more at low ray counts. The displayed $10^8$ point is excluded because it enters a different GPU memory and chunking regime. Every annotated method slope summarizes 12 independent macroimage fits.
    }
    \figurealttext{Three log-scale panels show raw-map error, source-convolved error, and runtime versus ray count. Errors decrease and runtime increases with more rays. IPM errors fall more steeply than IRS errors; the highlighted macroimage is compared with the other eleven cases.}
\label{fig:map_convergence_rate}
\end{figure*}

We fit all seven ray counts separately for each macroimage before taking the median of the 12 slopes. The raw fractional map RMSE gives $p=-0.84^{+0.03}_{-0.06}$, and source convolution steepens the median slope to $p=-0.87^{+0.07}_{-0.03}$. Q2237+0305 image B, shown prominently in Fig.~\ref{fig:map_convergence_rate}, reaches $p=-0.91\pm0.02$ for the raw map and $p=-0.90\pm0.02$ after source convolution. The quoted $\pm$ values for image B are individual-fit uncertainties, not the 12-macroimage percentile spread. The measured IPM slope near $-0.84$ is a finite-range rate over 1--100 million rays rather than an asymptotic prediction, with source-pixel intersections, caustic structure, the finite reference, and fixed reconstruction settings contributing to the measured error.

Although our internal IRS launches rays on a regular Cartesian lattice, its measured raw-map error has a median slope $p=-0.504^{+0.003}_{-0.003}$. This effectively reproduces the $N_{\rm ray}^{-1/2}$ counting-noise rate. The non-linear microlensing mapping largely disrupts the regular image-plane sampling before the rays are deposited into source pixels. Many source pixels receive few or no rays. Pixels receiving no rays are assigned zero magnification, without interpolation or an artificial ray count. These effects make the unsmoothed source-plane sampling behave more like random counting noise. Convolution with the extended source averages over neighbouring pixels and suppresses this small-scale ray-count noise. The remaining deterministic lattice and mapped-boundary error then becomes more important, steepening the median finite-range IRS slope to $-0.711^{+0.014}_{-0.036}$ and bringing it close to the expected Cartesian-lattice rate of $-3/4$. This change is specific to the regular lattice. GPU-D uses pseudorandom rays and therefore does not test this Cartesian-lattice behaviour. These empirical slopes depend on the tested lens-system sample.

\subsection{Method validation}
\label{subsec:method_validation}
\label{subsec:far_field_approx_results}
\label{subsec:scout_validation}

The fiducial local-exact/far-field configuration uses the order-ten Taylor scheme defined above. We fix the biquadratic IPM lattice and compare one complete map for each of the 12 macroimages with direct point-mass deflections, thereby isolating the deflection approximation. The audit also includes a conservative order-ten $16\times16$ nodal-grid, $\chi_{\rm near}=1.5$ setting, distinct from the $32\times32$ far-field grid used for the high-ray-count references. Across the 12 macroimages, the adopted $8\times8$, $\chi_{\rm near}=1$ calculation has a median fractional map RMSE of $3.1^{+2.6}_{-2.0}\times10^{-4}$ and a median magnitude-map RMSE of $0.34^{+0.30}_{-0.22}\,$mmag relative to direct point-mass deflections. The conservative control reduces these errors to $1.4^{+1.8}_{-0.80}\times10^{-4}$ and $0.15^{+0.21}_{-0.086}\,$mmag, respectively. Both metrics use the same definitions as Table~\ref{tab:accuracy_runtime}. 

The fast configuration estimates its normalization correction by comparing the adopted scout grid with a denser scout grid at frame zero (Section~\ref{subsec:lens_plane_scout}). The resulting uniform magnification offset is applied at subsequent epochs. The independent-frame controls omit this temporal correction, so batching alone does not explain the accuracy difference. The correction does not restore missing spatial structure or guarantee agreement with an independently evaluated denser scout. Increasing the determinant discovery grid from $512^2$ to $1024^2$ leaves the segment counts and source-centre labels unchanged.

The residuals show why the extent of the searched lens-plane region matters. In the representative Q2237+0305 image B map in Fig.~\ref{fig:method_comparison}, the Weisenbach residual has a mean of $+7.9\,$mmag and a centred spatial scatter of $4.1\,$mmag, compared with an uncentred RMSE of $8.9\,$mmag. The constant offset therefore contributes $79\,\mathrm{per\,cent}$ of the squared map error, and the mean fractional magnification deficit is $0.72\,\mathrm{per\,cent}$. The matched full-field control has only $-0.032\,$mmag bias, whereas the rectangular control has $+4.1\,$mmag bias. The complete light curves show the same qualitative signature. All 12 Weisenbach residuals have positive mean offsets, with a median bias of $6.2\,$mmag, a median centred scatter of $1.3\,$mmag, and a median RMSE of $6.3\,$mmag. The bias contributes a median $94\,\mathrm{per\,cent}$ of their squared error. We retain the uncentred errors in every benchmark rather than calibrating against the reference. These coherent offsets are consistent with finite shooting-field losses, although the residuals also include discretization and source-grid interpolation errors. Enlarging the Weisenbach shooting region would reduce the missing-flux contribution at additional cost if the ray density is maintained. A shared nominal light-loss tolerance therefore does not make a fixed rectangular shooting region equivalent to searching the full stellar aperture. Source scouting admits source-reaching cells from anywhere in that aperture, largely avoiding this coherent finite-field bias.

Automated numerical tests compare the uncompiled PyTorch, compiled PyTorch, and Triton implementations and check that temporal and independent-light-curve batching preserve fluxes and caustic labels. Selected elliptic-integral calculations are checked against SciPy~\citep{Scipy}, lens equations against NumPy~\citep{Numpy} and lenstronomy~\citep{BirrerAmara2018,Birrer2021Lenstronomy}, cosmological distances against Astropy~\citep{Astropy}, and bandpass integration against speclite~\citep{Kirkby2024Speclite}, complementing the map and caustic validations above and the SIM5 comparison in Section~\ref{subsec:gr_validation}.

\section{General-Relativistic Quasar Discs and Evolving Sources}
\label{sec:source_models}

Once magnification-map generation becomes sufficiently fast, source construction and CPU--GPU transfers can limit the speed of complete light-curve simulations. We therefore implement the evolving source models on the GPU, evaluating their emission and combining it with magnification maps in time and wavelength batches, with only the requested final products transferred to the CPU.

The microlensing calculation accepts any source represented on a fixed pixel grid. Here we couple it to a time-dependent, general-relativistic quasar disc with viscous heating, lamppost reprocessing, and thermal emission. The source conventions follow \citet{Fagin2025Reprocessing} and \citet{chan2024reverberationmappinglamppostwind}. As in \textsc{Amoeba} \citep{best2024amoebaagnmodeloptical}, we couple spatially resolved source emission, intrinsic variability, and microlensing. Our contribution is the GPU implementation of the ray-traced Kerr transfers and their batched coupling to dynamical microlensing. We describe this coupling below, with the disc-profile and geodesic details collected in Appendix~\ref{app:kerr_details}. The source model can also be used without microlensing for continuum reverberation calculations. The same interface accepts analytic profiles, supplied images, and time-ordered image cubes, as illustrated by the expanding supernova in Appendix~\ref{app:supernova}.

Each source image provides a nonnegative observer-frame flux density $F_{\nu,b,p}(t)$ in $\mathrm{erg\,s^{-1}\,cm^{-2}\,Hz^{-1}}$ for pixel $p$ and band $b$. Summing over pixels gives the unlensed flux. Weighting the pixels by their magnifications, with the appropriate macroimage delay, gives the lensed flux as in equation~\eqref{eq:finite_source_flux}. The fiducial quasar source uses a $1024^2$ grid enclosing 99.9 per cent of the modelled baseline flux, with a 5 per cent radial safety margin.

\subsection{Disc structure and intrinsic driving}
\label{subsec:disk_thermodynamics}
\label{subsec:quasar_variability}

We adopt a geometrically thin, optically thick equatorial disc with a zero-torque inner boundary at the innermost stable circular orbit (ISCO) and no emission at smaller radii \citep{Shakura_1973,Novikov_1973,PageThorne1974}. Writing $R_g=GM_{\rm BH}/c^2$, the viscous temperature is
\begin{equation}
 T_{\rm visc}^4(R)=
 \mathcal{N}_\beta
 \left(\frac{R}{R_g}\right)^{3-4\beta}
 \frac{3GM_{\rm BH}\dot M}{8\pi\sigma_{\rm SB}R_g^3}
 f_{\rm NT}(R,a),
\label{eq:viscous_temperature}
\end{equation}
where $\sigma_{\rm SB}$ is the Stefan--Boltzmann constant, $f_{\rm NT}(R,a)$ is the dimensionless Novikov--Thorne radial flux profile and $a$ is the signed spin. The factor $(R/R_g)^{3-4\beta}$ gives an asymptotic temperature slope $T_{\rm visc}\propto R^{-\beta}$, with $\beta=3/4$ recovering the standard thin disc \citep{Fagin2025Reprocessing}. The accretion rate is $\dot M=\lambda_{\rm Edd}L_{\rm Edd}/(\eta c^2)$, with $\eta$ the Novikov--Thorne radiative efficiency. The factor $\mathcal N_\beta$ preserves the disc-area integral of the local viscous flux, with $\mathcal N_{3/4}=1$. Appendix~\ref{app:disk_profile} gives the flux profile, efficiency, and explicit normalization.

The primary microlensing benchmarks use the mean disc brightness. For the variable-source demonstration, we generate a coronal driver with a smoothly broken power spectral density (PSD),
\begin{equation}
 P(\nu)=A\,\nu^{-\alpha_{\rm L}}
 \left[
 1+\left(\frac{\nu}{\nu_b}\right)^{
 \alpha_{\rm H}-\alpha_{\rm L}}
 \right]^{-1},
\label{eq:driver_psd}
\end{equation}
where $A$ sets the normalization, $\nu_b$ is the break frequency, and $\alpha_{\rm L}$ and $\alpha_{\rm H}$ are the low- and high-frequency slopes \citep{McHardy_2004,Uttley_2005}. Following \citet{Timmer_1995}, we draw Gaussian real and imaginary Fourier coefficients with variance proportional to $P(\nu)/2$, impose Hermitian symmetry, and Fourier-transform to the time domain. We generate five times the required duration and retain an interior segment to reduce periodic-boundary effects. After standardizing this latent sequence $Y(t)$, we ensure positive irradiation with
\begin{equation}
 \begin{gathered}
 X(t)=\exp\!\left[a_X+b_XY(t)\right],\\
 a_X=\ln\overline X-\frac{b_X^2}{2},\qquad
 b_X^2=\ln\!\left[1+\left(\frac{\sigma_X}{\overline X}\right)^2\right].
 \end{gathered}
\label{eq:positive_driver}
\end{equation}
For Gaussian $Y$, $\overline X$ and $\sigma_X$ are the target ensemble mean and standard deviation, not exact finite-sample moments. The PSD specifies $Y$, rather than the transformed driver $X$. We use $\overline X=1$ and an observer-frame grid with a 1000-day leading buffer for delayed evaluations. One driver illuminates all disc radii and wavelengths. Although lensing is linear in surface brightness, microlensing weights regions with different variability amplitudes and delays, so applying variability to the resolved source is generally not equivalent to modulating the final light curve.

\subsection{Relativistic observer and lamppost transfers}
\label{subsec:kerr_observer_transfer}
\label{subsec:lamppost_transfer}

The general-relativistic (GR) calculation traces a null geodesic from the centre of every source-screen pixel to its first equatorial-disc intersection \citep{Kerr_metric,Cunningham73}. We use analytic quartic roots and Jacobi elliptic functions for the intersection, and Carlson symmetric integrals for the emission coordinates and arrival time \citep{YangWang2013,GrallaLupsasca2020}. Appendix~\ref{app:kerr_separated} specifies the branches and numerical safeguards. The resulting static transfer records emission radius, azimuth, relative arrival time, and frequency shift
\begin{equation}
 g\equiv\frac{\nu_\infty}{\nu_{\rm em}}
 =\frac{1}{u^t(1-\Omega\lambda)},\qquad
 \Omega=\frac{1}{r^{3/2}+a},
\label{eq:kerr_gfactor}
\end{equation}
where $r=R/R_g$ on the equatorial disc, $\lambda$ is the photon angular-momentum constant, and $u^t$ is the circular-orbit four-velocity normalization \citep{Cunningham73,PageThorne1974,GatesHadarLupsasca2021}. We keep the disc orientation fixed, so positive and negative spin describe prograde and retrograde orbits relative to that orientation. The model retains only the direct primary image and evaluates brightness at the actual emission coordinates without an oversampled screen or ray-deposition kernel.

We adopt two identical, isotropically emitting lampposts at $\pm H_{\rm lamp}$, each with X-ray luminosity $L_X$. By symmetry, we trace the upper source and include both sources in the heating normalization. With albedo $A_{\rm alb}$, the corresponding Newtonian profile is \citep{chan2024reverberationmappinglamppostwind}
\begin{equation}
 F_{\rm lamp}^{\rm flat}(R)=
 \frac{(1-A_{\rm alb})L_X}{2\pi}
 \frac{H_{\rm lamp}}{(R^2+H_{\rm lamp}^2)^{3/2}}.
 \label{eq:flat_lamppost_flux}
\end{equation}
The configured height is $H_{\rm lamp}/R_g=\epsilon+R_{\rm ISCO}/R_g$. The relativistic calculation follows locally emitted rays to their first disc intersection and constructs
\begin{equation}
 {\cal K}^{\rm raw}_{\rm lamp}(R)=
 g_{\rm ld}^{p_g}(R)\frac{d\Omega}{dA_{\rm proper}},
\label{eq:gr_lamppost_kernel}
\end{equation}
where $g_{\rm ld}$ is the lamp-to-disc frequency shift and $dA_{\rm proper}$ is the receiving disc-comoving area in units of $R_g^2$. We use $p_g=2$ for the photon-energy and arrival-rate factors \citep{WilkinsFabian2012}. Equation~\eqref{eq:app_lamp_normalization} normalizes this kernel to the proper-area integral of the flat profile, retaining relativistic redistribution without changing the integrated heating. The comoving area and finite-axis ray calculation are specified in Appendix~\ref{app:kerr_lamppost} \citep{Dauser2013,Ingram2019}.

The two transfers give the observer-frame relative delay
\begin{equation}
 \tau(R,\phi)=
 (1+z_s)\frac{R_g}{c}
 \left[
 t_{\rm ld}(R)+t_{\rm do}(R,\phi)
 -t_{\rm min}
 \right].
\label{eq:total_disk_delay}
\end{equation}
Here $t_{\rm ld}$ and $t_{\rm do}$ are lamp-to-disc and disc-to-observer coordinate times in gravitational units. We subtract their minimum sum over valid pixels because only relative lags are needed, making the earliest retained disc response occur at $\tau=0$. The factor $(1+z_s)R_g/c$ converts to observer-frame seconds, then to days for the driver. Comparison with a directly observed coronal signal would instead require the lamp-to-observer arrival time as the reference.

\subsection{Time-dependent observer-frame emission}
\label{subsec:observer_brightness}

The normalized lamp kernel sets
\begin{equation}
 T_{\rm lamp}^4(R)=
 \frac{f_{\rm lamp}\eta}{\lambda_{\rm Edd}}
 \frac{3c^2\dot M}{8\pi\sigma_{\rm SB}R_g^2}\,
 \mathcal{K}_{\rm lamp}(R),
 \label{eq:lamp_temperature}
\end{equation}
and the evolving temperature is
\begin{equation}
 T_{\rm eff}^4(R,\phi,t)=T_{\rm visc}^4(R)
 +T_{\rm lamp}^4(R)\,
 X\!\left[t-\tau(R,\phi)\right].
\label{eq:time_dependent_temperature}
\end{equation}
The local bolometric flux is $F_{\rm bol}=\sigma_{\rm SB}T_{\rm eff}^4$. Following \citet{Fagin2025Reprocessing}, $f_{\rm lamp}\equiv(1-A_{\rm alb})L_X/L_{\rm Edd}$ sets the mean irradiation and is partly degenerate with $\sigma_X$ in the variable response.

Each disc element emits a colour-corrected blackbody with $T_{\rm col}=f_{\rm col}T_{\rm eff}$ and dilution $f_{\rm col}^{-4}$ \citep{DavisElAbd2019,chan2024reverberationmappinglamppostwind}. Conservation of $I_\nu/\nu^3$ gives $I_{\nu,\infty}=g^3 I_{\nu/g,{\rm em}}$, evaluated at the quasar-rest-frame wavelength $\lambda_{b,\infty}=\lambda_{b,{\rm obs}}/(1+z_s)$. Cosmological dimming then gives $I_{\nu,b}^{\rm obs}=(1+z_s)^{-3}I_{\nu,\infty}$. For physical screen spacing $\Delta x_p$, the observed pixel flux is $F_{\nu,b,p}=I_{\nu,b,p}^{\rm obs}(\Delta x_p/D_A)^2$. Equivalently, $F_{\nu,b,p}=(1+z_s)I_{\nu,\infty,p}(\Delta x_p/D_L)^2$, where $D_L=(1+z_s)^2D_A$. The explicit blackbody expression is given in Appendix~\ref{app:source_observables}. Pixels with no valid primary disc intersection contribute zero. The disc position angle rotates the observer screen relative to the microlensing field.

We use $f_{\rm col}=1$ and evaluate each band at its observed effective wavelength, rather than integrate over the filter throughput. The adopted $ugrizy$ wavelengths are $(3671,4827,6223,7546,8691,9712)\,$\AA. The reddest rest-frame profile determines the source extent. The Kerr and lamppost transfers are calculated once for each geometry and reused across times and wavelength batches. At each source-evaluation time, we update the delayed driver, temperature, and thermal emission, then integrate the brightness against the magnification map.

\subsection{Continuum transfer functions}
\label{subsec:continuum_transfer_functions}

For small variations around the mean driver, define the local response weight $W_b(\boldsymbol\beta)=\left.\partial I_{\nu,b}^{\rm obs}/\partial X\right|_{X=1}$. Its explicit form is given in Appendix~\ref{app:source_observables}. The microlensed response is
\begin{equation}
 \psi_{i,b}(\tau;t)=
 \int_{\Omega_\beta}\mu_i(\boldsymbol\beta,t)
 W_b(\boldsymbol\beta)
 \delta\!\left[\tau-\tau(\boldsymbol\beta)\right]
 d^2\boldsymbol\beta,
\label{eq:microlensed_response}
\end{equation}
where $\delta$ is the Dirac delta distribution and $\tau(\boldsymbol\beta)$ is the disc delay. The corresponding mean lag is
\begin{equation}
 \langle\tau\rangle_{i,b}(t)=
 \frac{\int \tau\psi_{i,b}(\tau;t)\,d\tau}
      {\int \psi_{i,b}(\tau;t)\,d\tau}.
\label{eq:microlensed_mean_delay}
\end{equation}
Setting $\mu=1$ gives the intrinsic response. Differential magnification reweights regions with different light-travel times, changing the response shape and mean lag \citep{TieKochanek2018,Chan_2020,Fagin2025Reprocessing}. This image-dependent response acts in addition to the macroimage delay and can bias inter-image delay measurements if neglected \citep{TieKochanek2018}. In the Fig.~\ref{fig:end_to_end_hst_lsst} snapshot, image A's $u$-band mean response lag decreases from $5.4$ to $4.4$ days, a shift of $-0.98$ days. This is a response-centroid shift, not a recovered cosmological-delay error.

The light curves reported here use the full non-linear source emission rather than the response approximation described in Appendix~\ref{app:source_observables}.

\subsection{Multi-rate intrinsic variability and microlensing}
\label{subsec:multirate_temporal}

Intrinsic source variability can be quickly evaluated more frequently than the evolving microlensing maps. Let $\mathcal{M}_{i,j}(\boldsymbol{\beta})$ denote the magnification map of macroimage $i$ at time $t_j$ on a fixed source-plane grid. For a source-evaluation time $t_j\leq t\leq t_{j+1}$, define $w=(t-t_j)/(t_{j+1}-t_j)$. The flux in band $b$ is approximated by
\begin{equation}
    \begin{split}
    F_{\nu,i,b}^{\,\rm L}(t) \simeq {}&
    \int d^2\boldsymbol{\beta}\,
    I_{\nu,b}\!\left(\boldsymbol{\beta},t-\Delta t_i\right)\\
    &{}\times\left[
        (1-w)\mathcal{M}_{i,j}(\boldsymbol{\beta})
        + w\mathcal{M}_{i,j+1}(\boldsymbol{\beta})
    \right],
    \end{split}
    \label{eq:multirate_map_interpolation}
\end{equation}
where $I_{\nu,b}$ is the evolving observer-frame source brightness and $\Delta t_i$ is the macroimage time delay, as in equation~\eqref{eq:finite_source_flux}. At each requested time, we calculate the flux using each neighbouring map on the same pixel grid and linearly combine the two fluxes. This requires no spatial resampling or intermediate map. Evaluating the source at its own cadence retains changes in its spatial profile that scalar light-curve interpolation would miss. Between map epochs, interpolation blends neighbouring magnification maps rather than following individual caustics, so rapidly changing patterns may require a shorter map cadence. For example, we can evaluate the source and light curve daily while calculating microlensing maps every 25 days. Across the 12 benchmark macroimages, evaluating the variable Kerr source daily while interpolating 25-day magnification maps differs from direct daily-map light curves by a median six-band RMSE of $0.28^{+0.45}_{-0.25}\,$mmag.

\subsection{End-to-end LSST-like observations}
\label{subsec:hst_lsst}

\begin{figure*}
\centering
\includegraphics[width=\linewidth]{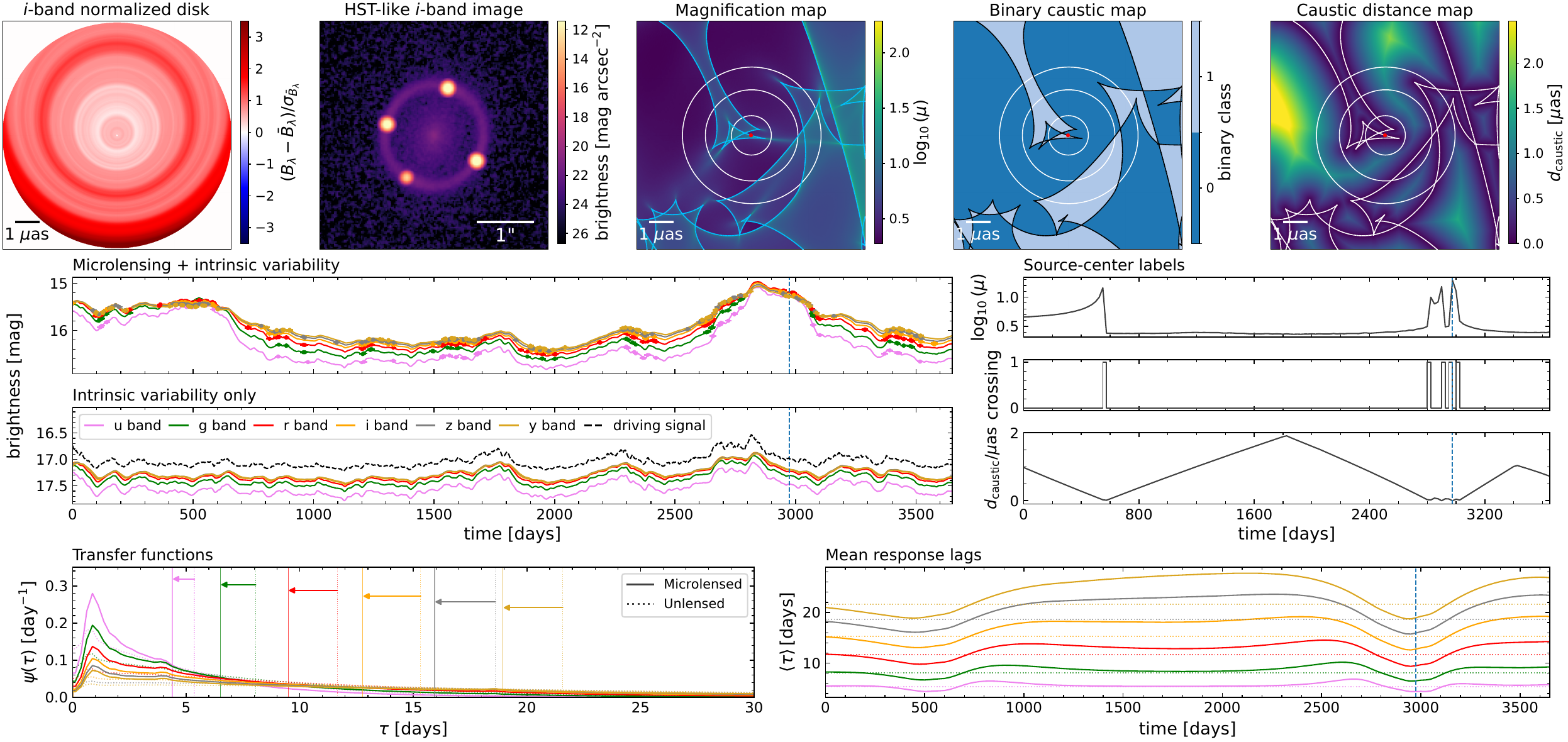}
\caption{
End-to-end Q2237+0305-like simulation at $t=2975$ days, showing macroimage A. The top row shows a pixelwise standardized $i$-band disc fluctuation, the mock HST-like lens image, magnification, binary caustic class, and caustic distance. The middle row compares the combined and intrinsic light curves alongside the source-centre histories. Mock WFD observations appear only on the combined light curve, and the dashed black curve shows the rescaled driver. The bottom row shows response functions at 0.25-day spacing and mean lags, with dotted unlensed comparisons and arrows indicating the microlensing-induced lag shifts. The plotted responses $\psi$ are normalized to unit area. The light curves and response functions use the same $1024^2$ source grid. Source emission is evaluated daily and maps every 25~days, with $N_{\rm ray}=10^7$, $k=2$, $r=2$, and $v=4$.
}
\figurealttext{A multi-panel example connects a fluctuating quasar disc and mock lens image to magnification and caustic maps, multiband light curves, source-centre labels, response functions, and mean lags. Arrows compare microlensed and unlensed mean response delays.}
\label{fig:end_to_end_hst_lsst}
\end{figure*}

We demonstrate the complete pipeline with a four-image Q2237+0305-like lens. Each macroimage has its own lens parameters and independently seeded stellar field, while the shared source is evaluated at the appropriate macroimage arrival time. Fig.~\ref{fig:end_to_end_hst_lsst} shows image A's maps, caustic labels, response functions, and light curves. The driver has $(\alpha_{\rm L},\alpha_{\rm H})=(1,3)$, $1/\nu_b=100$ days, and $\sigma_X=0.3$. Daily source emission is combined with the neighbouring 25-day maps using equation~\eqref{eq:multirate_map_interpolation}.

The mock lens image uses the PyTorch-native \textsc{caustics} package \citep{Stone2024Caustics}, with an elliptical power-law lens, external shear, extended lens and host light, and variable quasar images \citep{TessoreMetcalf2015,Shajib2024StrongLensing}. It is an HST-like visualization rather than a calibrated reproduction of an observation. We sample the light curves at the times and filters of 869 visits from a WFD field in the Rubin \texttt{baseline\_v4.3.5\_10yrs} OpSim realization.\footnote{\url{https://rubin-sim.lsst.io/data-download.html}.} The visit depths set the photometric noise through the standard LSST prescription, with a 0.005 mag systematic floor \citep{LSST,Fagin2025Reprocessing}. Rendering and noise parameters are given in Appendix~\ref{app:mock_observations}. These samples do not include image-level deblending of the macroimages. Extracting such curves from survey images requires scene modelling, for example with STARRED \citep{Millon2024STARRED}.

An optional package spectral layer supplements the physical disc continuum with empirical emission lines, dust, and intergalactic absorption before bandpass integration \citep{Temple2021QuasarColours,Fagin2025Reprocessing}. Lines receive macro magnification but no stellar microlensing, and host light requires an explicit lensing prescription. The figures and benchmarks here instead use the six effective-wavelength disc calculations.

\subsection{General-relativistic validation and performance}
\label{subsec:gr_validation}

For the direct primary observer-to-disc transfer tested here, the analytic Kerr calculation reaches exactly the same valid observer-screen pixels as SIM5 and gives image normalized mean-square errors of $6.2\times10^{-11}$--$1.2\times10^{-8}$ for the three fiducial systems. After warm-up, the primary-image transfer takes $1.8$--$1.9\,$ms, compared with $0.54$--$0.57\,$s for the matched single-threaded SIM5 calculations. The detailed image and residual comparison is shown in Appendix~\ref{app:kerr_implementation}.

The matched SIM5 comparison isolates only the primary observer-to-disc ray trace and therefore does not characterize the cost of constructing the full relativistic source. Profiling of the benchmark implementation after compilation and warm-up gives a median $0.081\,$s to construct the complete $1024^2$ disc source across the three benchmark prescriptions. This calculation includes the primary image, relative observer delays, lamppost illumination and delay transfer, disc normalization, and intrinsic driving realization. The observer and lamppost transfers are calculated once for a fixed disc geometry and reused throughout the light curve. The one-time compilation cost is excluded. Evaluating the complete 147-epoch, six-band surface-brightness sequence in temporal batches of 49 takes a median $0.069\,$s. Each source prescription is shared by the four macroimages of its system. These source costs are reported separately from the microlensing-method timings. Rubin visit selection and photometric-noise sampling are also excluded because they are shared post-processing operations rather than part of the magnification calculation.

\subsection{Other time-dependent sources}
\label{subsec:alternative_sources}

An expanding Type~Ia-like photosphere demonstrates the same interface with changing luminosity, colour, and spatial extent. We use an early $L\propto t^{2.4}$ rise \citep{Firth2015}, an exponential decline, and homologous expansion. The projected elliptical, limb-darkened profile is initially achromatic, motivated by \citet{Goldstein2018SN}, with weak chromatic structure after 21 rest-frame days. Compact early emission is sensitive to fine caustic structure, while expansion averages over it and the later band-dependent spatial profiles allow chromatic microlensing. This is a source-interface demonstration rather than a precision supernova radiative-transfer model. Its prescription and matched 12-macroimage benchmark are given in Appendix~\ref{app:supernova}.

The benchmark evaluates the source and maps at 51 epochs over 200 observer-frame days. The fast method gives a median light-curve RMSE of $0.44\,$mmag and runtime of $0.14\,$s. A separate test with daily source evaluation finds that four-day maps differ from direct daily-map light curves by a median six-band RMSE of $0.056^{+0.21}_{-0.029}\,$mmag. Finer map cadences may be needed near rapidly evolving caustics.

\section{Discussion}
\label{sec:discussion}

\subsection{Origins of the speedup and relation to prior methods}
\label{subsec:speedup_interpretation}

The speedup is cumulative rather than attributable to one approximation, and most of its ingredients build on established ideas. Our comparisons therefore test complete implementations under a common source and stellar field rather than claim that each numerical idea is new in isolation. The fast source-scouted path is $19$ times faster than the batched full-field IPM control at nearly the same fractional map RMSE. Its calculation of one light curve is about $710$ times faster than the compared Weisenbach calculation, while batching independent curves raises the measured throughput advantage to $780$ times.

Tree methods group distant microlenses hierarchically \citep{Wambsganss1990,Bate2010}. \citet{Zheng2022GPU} combine exact nearby deflections with an interpolated far field, while FMM-based methods use multipole and local expansions \citep{JimenezVicente2022,Weisenbach2025Maps}. Our local-exact/far-field solver follows the same near/far principle but uses regular cell-local expansions that can be shared across rays and constructed for several epochs at once. For spatial selection, \citet{Meena2022AdaptiveBoundary} showed how image-plane pixels near a compact source can be refined adaptively. Our scout searches the full circular stellar aperture and retains cells whose mapped footprints reach the requested source field. It avoids most noncontributing area while retaining source-reaching cells that a fixed rectangular window can miss. Between scout evaluations, we retain cells selected at either endpoint. The frame-zero correction removes the dominant normalization offset of the faster $k=2$ selection.

The area-conserving principle of IPM was introduced and optimized by \citet{Mediavilla2006,Mediavilla2011}. Virtual refinement resolves the biquadratic mapping more finely without additional lens-equation evaluations. Fused Triton rasterization then avoids the large intermediate tensors required by the corresponding PyTorch calculation. Shared ray tracing, interpolation, and rasterization remain the dominant cost in the stage profiles reported in Section~\ref{subsec:qualitative_results}. The IPM maps show a median raw-map convergence slope at $-0.84$, whereas both our Cartesian IRS and the pseudorandom GPU-D calculation scale close to $N_{\rm ray}^{-1/2}$ in these caustic-rich fields.

Moving-star calculations showed that evolving caustics can produce shorter, steeper, and more frequent events than bulk motion alone \citep{WambsganssKundic1995}. Processing several epochs together allows fixed-size arrays and allocated GPU memory to be reused. Batching independent light curves further increases throughput. For caustic labels, \citet{Weisenbach2025} traces the complete critical-curve network and describes winding-number labels from oriented caustics. We instead evaluate $\det\mathbf{J}$ and apply marching squares only within the scouted region. Anchors and gauges then track the binary caustic class at the source centre and its distance to the nearest caustic without a complete winding map at every epoch. Caustic construction remains the largest additional cost of the labelled workload.

The adopted configuration is deliberately conservative, with sub-millimagnitude median light-curve error and sub-percent median raw fractional map RMSE. Less stringent applications can reduce the ray density or spatial refinement, while light-curve-only calculations can omit the determinant and labelling passes. The lens and source calculations remain on the GPU until final products are transferred to the CPU. PyTorch and Triton's support for NVIDIA and AMD backends also offers portability beyond CUDA-only implementations.

\subsection{Relativistic source effects}
\label{subsec:gr_science}
Relativistic frequency shifts make the observed disc brightness asymmetric even for an axisymmetric rest-frame temperature profile. Spin sets the ISCO and changes the innermost emitting area, while inclination strengthens the projected asymmetry and Doppler contrast. During a caustic crossing, the high-magnification region scans the Doppler-brightened side, strongly redshifted inner disc, and regions with different reverberation delays. It therefore changes the band-dependent transfer function and mean lag rather than merely rescaling the total flux.

\subsection{Limitations}
\label{subsec:limitations}

The benchmark uses three lenses, four macroimages per lens, and one stellar realization per macroimage, with fixed source and population prescriptions. It is a controlled comparison of numerical methods, not a representative population of lensed quasars. Each additional realization requires a high-quality reference and runs of the much slower external comparison codes, limiting the size of this matched benchmark. The distribution of runtime and error in a diverse training set requires a broader study. The primary quasar timings measure 147 microlensing epochs separated by 25 days. They do not include construction of the GR disc, daily intrinsic variability, or the LSST-like sampling and noise model. These shared source and observation stages are excluded so the benchmark isolates the microlensing methods.

The timings are specific to the RTX 5070 Ti, which is a moderate consumer GPU. Batch and chunk sizes should be retuned for the available hardware. The simulator can use the larger memory and parallel capacity of high-performance GPUs through temporal batching and batching across independent light curves. These forms of batching are not available in the comparison implementations, so larger GPUs can provide an additional practical advantage for our method. Our simulator also works on CPU, but will be significantly slower.

The source scout traces only lens-plane cells selected as capable of reaching the finite source. The initial correction reduces the dominant magnification offset in the faster $k=2$ selection, but finite support and finite ray density can still omit a small amount of flux or an unusually small caustic component. Increasing the ray count, scout support, or determinant resolution reduces these risks at additional cost. The adopted $8192^2$ determinant field can also miss caustics below its resolved scale.

The anchor and gauge method does not mathematically guarantee that every caustic crossing will be identified. The adopted nine anchors and nine gauges provide strong redundancy, and no reference crossing was missed in the 12 benchmark light curves. However, very dense caustics, incomplete components, or large changes in the caustic pattern between successive frames can challenge the classification. More anchor and gauge points and shorter temporal intervals increase robustness while raising the runtime. Likewise, temporal scout reuse assumes that the stellar field evolves slowly relative to the refresh interval. This condition is satisfied in the tested systems but should be validated for substantially larger velocities or longer cadences.

The source grid encloses nearly all modelled flux and resolves the wavelength dependent disc, but its fixed $1024^2$ sampling only marginally resolves the smallest inner-disc scales in some configurations. Adaptive or locally refined source grids would better isolate the ISCO and compact caustic-crossing structure. The calculations use primarily float32 arithmetic, which is sufficient for the reported map, light-curve, and GR validation errors. Float64 remains available when stricter numerical precision is required, with a substantial cost in memory and runtime.

The benchmark quasar source assumes a thin equatorial disc, symmetric point lampposts, and the primary Kerr image. The optional empirical spectral layer adds unresolved line and attenuation components, but neither it nor the benchmark source provides a self-consistent, spatially resolved broad-line region, diffuse continuum, or extended corona. The supernova uses effective band wavelengths and a simplified photosphere rather than detailed radiative transfer. Users can replace either source through the common image interface. Linear interpolation between dynamical maps can also blur rapidly moving caustics. Applications requiring stricter temporal accuracy should insert intermediate maps and validate the cadence for their velocities and source size.

\subsection{Applications and future work}
\label{subsec:outlook}
\label{subsec:source_generality}

The immediate application is LSST-scale forward modelling in which many independent dynamical microlensing histories are combined with source variability and sparse survey sampling. Future work will use matched static and dynamical calculations to determine how neglecting stellar motion biases inferred quasar disc sizes. Large independent ensembles will also support simulation-based inference of quasar parameters and cosmological time delays, as well as predictions of caustic crossings and HMEs for follow-up \citep{Vernardos_2019,fagin2024predictinghighmagnificationevents}. The multi-rate interface is important for this scaling because a daily source light curve does not require a daily stellar-field ray trace once the map cadence has been validated.

Future implementations could allocate resolution adaptively in both time and space, inserting microlensing epochs when caustic-distance diagnostics indicate an approaching crossing and locally refining source and magnification-map sampling around compact inner-disc emission and nearby caustic structure.

The expanding-supernova example demonstrates evolving luminosity, colour, and source extent through the same image interface. Detailed radiative-transfer profiles and non-spherical ejecta could replace the prototype without changing the map solver. Other applications include ring-like optical gamma-ray-burst afterglows \citep{LoebPerna1998,Garnavich2000GRB}, kilonovae, and tidal-disruption flares, provided suitable evolving brightness profiles are supplied \citep{ChenLuChen2024LensedTDE,GangulyMore2026LensedKilonovae}.

Natural source extensions include continuum and velocity-resolved reverberation, extended coronal geometries, Fe K$\alpha$ emission, X-ray microlensing, broad-line regions, and diffuse continuum emission \citep{Morgan_2012,Paic2022,best2024amoebaagnmodeloptical}. The same numerical pipeline can study planetary and compact-dark-matter microlenses in macrolensed quasars \citep{Mediavilla2017PBH,DaiGuerras2018,Tuntsov2024Planets}. Compound microlensing in double-source-plane lenses is another possible extension, but stars in multiple deflector planes would require a multi-plane version of the present single-plane calculation \citep{Salama2026DoubleSourcePlane}.

Galactic stellar and planetary microlensing is another extension. Explicit point-mass positions and masses, source trajectories, and pixelated stellar brightness profiles fit the present interfaces. A complete exoplanet-event model would need event-specific geometry, limb darkening, blending, parallax, and orbital lens trajectories where relevant, followed by validation against established planetary calculations \citep{Gaudi2008Planets,Penny2011OrbitalMotion}.

Cluster-scale macrocaustics provide another application. Intracluster stars break them into microcaustic networks that can repeatedly magnify individual background stars \citep{Venumadhav2017ClusterCaustics,Kelly2018Icarus}. Applying the present local model in this extreme macrocritical regime will require dedicated validation.

The fast evaluation of evolving point-mass lens fields might also inform gravitational-wave and fast-radio-burst microlensing \citep{Munoz2016FRBLensing,Diego2019GWMicrolensing}. Predicting waveform interference or delayed radio pulses, however, requires arrival-time or wave-optics calculations beyond the present intensity-map model.

\section{Conclusions}
\label{sec:conclusions}

We have presented \textsc{Microcaustics}, a GPU-accelerated
simulator for dynamical microlensing of time-dependent extended
sources. The method combines a local-exact/far-field deflection
solver, source-directed lens-plane scouting, biquadratic IPM,
fused Triton rasterization, temporal batching, and reuse of
scouted cells. Selective mapping and fused rasterization also accelerate
static maps, while temporal batching and reuse provide additional
gains for dynamical sequences.

Across the 12-macroimage quasar benchmark, the adopted configuration
reaches a median light-curve RMSE of $0.35\,\mathrm{mmag}$ and a raw
fractional map RMSE of $7.2\times10^{-3}$ relative to the
high-resolution reference, recovering all 41 reference
source-centred caustic crossings. For the ten-year, 147-epoch,
six-band microlensing calculation on an RTX 5070 Ti GPU, concurrent
evaluation gives median amortized runtimes of $0.68\,\mathrm{s}$
per macroimage, or $0.88\,\mathrm{s}$ with caustics and labels.
These timings isolate the microlensing calculation after
compilation warm-up. The light-curve-only throughput is
approximately 780 times higher than that of the compared CUDA IPM
implementation at comparable map accuracy, and approximately 2500 times faster including caustic crossing labels. Convergence tests give
a median effective raw-map scaling of
$N_{\rm ray}^{-0.84}$, approaching the idealized
$N_{\rm ray}^{-1}$ piecewise-affine IPM rate.

The same GPU-resident framework includes a relativistic quasar
source that couples stochastic driving variability to spatially
resolved disc emission and reverberation. Its analytic primary
Kerr transfer agrees with SIM5 and is approximately 300 times
faster than the matched single-threaded SIM5 calculation after
compilation warm-up. End-to-end examples combine intrinsic
variability, independent macroimage microlensing, macroimage
time delays, microlensed response functions, and LSST-like
sampling. An expanding-supernova prototype demonstrates the
generic-source interface, producing 51-epoch light curves with
a median RMSE of $0.44\,\mathrm{mmag}$ and a median amortized
runtime of $0.12\,\mathrm{s}$ per curve.

Together, these results change dynamical microlensing from an
exceptional calculation into a scalable component of forward
modelling. Generating evolving stellar fields, time-dependent
relativistic sources, and caustic-crossing information within
one framework provides a foundation for Bayesian and
simulation-based inference of quasar accretion-disc structure, microlensing-induced time-delay biases for time-delay cosmography, and forecasting of caustic crossings and high-magnification events.
\textsc{Microcaustics} is open-source and available at
\url{https://github.com/JFagin/microcaustics}.

\section*{Acknowledgements}
Support was provided by Schmidt Sciences, LLC, for J.F., J.C., S.M., M.O., L.P.L., and Y.H\@. S.B. is supported by the US Department of Energy Grant DE-SC0026113.
C.S. acknowledges the support of an NSERC Postdoctoral Fellowship and a CITA National Fellowship.

This work made use of ChatGPT and Codex (\mbox{OpenAI}) assistance for code development and manuscript editing. The authors reviewed and take responsibility for all the content of this paper.

\noindent Software used: \textsc{PyTorch}~\citep{Pytorch}, \textsc{Triton}~\citep{Tillet2019Triton}, \textsc{caustics}~\citep{Stone2024Caustics}, \textsc{Matplotlib}~\citep{Matplotlib}, \textsc{NumPy}~\citep{Numpy}, \textsc{SciPy}~\citep{Scipy}, \textsc{Astropy}~\citep{Astropy}, \textsc{lenstronomy}~\citep{BirrerAmara2018,Birrer2021Lenstronomy}, \textsc{speclite}~\citep{Kirkby2024Speclite}, \textsc{SIM5}~\citep{Sim5_2}, Weisenbach 2025~\citep{Weisenbach2025Maps,Weisenbach2025}, GPU-D~\citep{VernardosFluke2014Cloud}.

\section*{Data availability}

The \textsc{Microcaustics} software is publicly available at \url{https://github.com/JFagin/microcaustics}. The benchmark data underlying this article are available from the corresponding author upon reasonable request.

\bibliographystyle{mnras}
\bibliography{bib}

\onecolumn
\appendix
\numberwithin{equation}{section}
\renewcommand{\theequation}{\Alph{section}\arabic{equation}}
\renewcommand{\theHequation}{\Alph{section}.\arabic{equation}}

\section{Disc Source and Relativistic Transfer Details}
\label{app:kerr_details}

This appendix specifies the source prescription and numerical choices used in Section~\ref{sec:source_models}. The calculation has three stages. The disc profile fixes viscous heating. The observer and lamppost transfers then assign emission coordinates, irradiation, frequency shifts, and delays to the fixed source grid. Finally, the delayed driver determines the brightness and response weights. The transfers are static for a given geometry and are reused throughout a light curve.

\subsection{Radial disc profile and normalization}
\label{app:disk_profile}

We use $R_g=GM_{\rm BH}/c^2$, signed spin $-0.998\leq a\leq0.998$, and the equatorial ISCO radius of \citet{BardeenPressTeukolsky1972}, with $R_{\rm ISCO}=6R_g$ at $a=0$. Positive and negative spin describe prograde and retrograde motion relative to the fixed disc orientation. The Eddington luminosity is $L_{\rm Edd}=4\pi GM_{\rm BH}m_pc/\sigma_T$, where $m_p$ is the proton mass and $\sigma_T$ the Thomson cross section. The adopted efficiency is $\eta_{\rm NT}=1-[1-2R_g/(3R_{\rm ISCO})]^{1/2}$, compared with the Newtonian value $\eta_{\rm SS}=R_g/(2R_{\rm ISCO})$.

For completeness, the dimensionless Novikov--Thorne profile used in equation~\eqref{eq:viscous_temperature} is \citep{PageThorne1974}
\begin{equation}
 \begin{split}
 f_{\rm NT}(R,a)={}&
 \frac{1}{x^7-3x^5+2ax^4}
 \Bigg[x-x_0-\frac{3a}{2}\ln\!\left(\frac{x}{x_0}\right)
 -\frac{3(x_1-a)^2}{x_1(x_1-x_2)(x_1-x_3)}
 \ln\!\left(\frac{x-x_1}{x_0-x_1}\right)\\
 &\hphantom{\frac{1}{x^7-3x^5+2ax^4}\Bigg[}
 {}-\frac{3(x_2-a)^2}{x_2(x_2-x_1)(x_2-x_3)}
 \ln\!\left(\frac{x-x_2}{x_0-x_2}\right)
 -\frac{3(x_3-a)^2}{x_3(x_3-x_1)(x_3-x_2)}
 \ln\!\left(\frac{x-x_3}{x_0-x_3}\right)\Bigg],
 \end{split}
\label{eq:novikov_thorne_flux_factor}
\end{equation}
where $x=(R/R_g)^{1/2}$, $x_0=(R_{\rm ISCO}/R_g)^{1/2}$, $x_1=2\cos[\tfrac13\cos^{-1}a-\pi/3]$, $x_2=2\cos[\tfrac13\cos^{-1}a+\pi/3]$, and $x_3=-2\cos(\tfrac13\cos^{-1}a)$. The corresponding Shakura--Sunyaev expression is $f_{\rm SS}=(R_g/R)^3[1-(R_{\rm ISCO}/R)^{1/2}]$ \citep{Shakura_1973}.

The normalization in equation~\eqref{eq:viscous_temperature} preserves the disc-area integral of the local viscous flux:
\begin{equation}
 \mathcal{N}_\beta =
 \frac{\int_{R_{\rm ISCO}}^{R_{\rm out}}
 f_{\rm NT}(R,a)\,R\,dR}
 {\int_{R_{\rm ISCO}}^{R_{\rm out}}
 f_{\rm NT}(R,a)(R/R_g)^{3-4\beta}\,R\,dR}.
\label{eq:viscous_power_normalization}
\end{equation}
We evaluate these integrals on 20,000 logarithmically spaced radii up to a fixed $R_{\rm out}=10^6R_g$. This normalization domain is distinct from the source-image extent, which is chosen from the reddest baseline monochromatic profile. The infinite-radius normalization converges only for $\beta>1/2$, but the finite domain remains defined for every supported $\beta$.

\subsection{Observer screen to primary disc intersection}
\label{app:kerr_separated}

For each observer-screen pixel, we trace a photon backward to its first equatorial disc intersection to find its emission radius and frequency shift. SIM5 inspired this implementation \citep{Sim5,Sim5_2}. The equations below specify our conventions and branch choices within established analytic Kerr solutions. We use Boyer--Lindquist coordinates $(r,\theta,\phi)$ with $G=M_{\rm BH}=c=1$. Radius is measured in units of $R_g$, with $r=R/R_g$ on the disc, and times are recovered using $R_g/c$.

The screen position fixes the conserved quantities that determine the ray's trajectory. For inclination $i$ and Bardeen screen coordinates $(\alpha,\beta_{\rm im})$, these are \citep{Cunningham73,YangWang2013}
\begin{equation}
 \lambda=-\alpha\sin i,\qquad
 q^2=\beta_{\rm im}^2+
 (\alpha^2-a^2)\cos^2 i,
\label{eq:kerr_screen_constants}
\end{equation}
where $\lambda=L_z/E$ and $q^2=Q/E^2$ normalize the axial angular momentum and Carter constant by the photon energy at infinity. Carter's separation of the geodesic equations allows the radial and polar motion to be solved separately \citep{Carter1968,GrallaLupsasca2020}. Their allowed ranges and turning points are determined by the potentials, with $\Delta=r^2-2r+a^2$,
\begin{equation}
 {\cal R}(r)=(r^2+a^2-a\lambda)^2-\Delta\left[q^2+(\lambda-a)^2\right],
 \qquad\quad
 \Theta(\theta)=q^2+a^2\cos^2\theta-\lambda^2\cot^2\theta.
\label{eq:kerr_potentials}
\end{equation}
Using Mino time $\gamma_{\rm M}$, the motion satisfies $(dr/d\gamma_{\rm M})^2=\mathcal R$ and $(d\theta/d\gamma_{\rm M})^2=\Theta$ \citep{Mino2003}. This is a path parameter, not the photon travel time. The polar solution first gives the observer-to-equator interval $\gamma_d$ at which the ray reaches $\theta=\pi/2$.

Substituting $u=\cos^2\theta$ into $\Theta=0$ gives a quadratic whose roots determine the polar turning points. To express the polar integral in elliptic form, we use
\begin{equation}
 B=a^2-q^2-\lambda^2,\qquad
 D=(B^2+4a^2q^2)^{1/2},\qquad
 u_+=\frac{2q^2}{D-B},\qquad
 A_-=\frac{B-D}{2}.
\label{eq:app_polar_roots}
\end{equation}
Here $u_+$ gives $\cos^2\theta$ at the polar turning point. Following equations (29)--(35) of \citet{GrallaLupsasca2020}, define $m_\theta=a^2u_+/A_-\leq0$, $c_\theta=|A_-|^{-1/2}$, and $s_o=\cos i/\sqrt{u_+}$, with $m_\theta=0$ in the nonspinning limit. For an asymptotic observer at $\theta_o=i<\pi/2$, the parameter interval to the first disc crossing is
\begin{equation}
 \gamma_d=c_\theta
 \begin{cases}
  F(\sin^{-1}s_o\mid m_\theta), & \beta_{\rm im}\leq0,\\
  2K(m_\theta)-F(\sin^{-1}s_o\mid m_\theta),
    & \beta_{\rm im}>0
 \end{cases}.
\label{eq:app_polar_mino}
\end{equation}
We write the incomplete first-kind elliptic integral as $F(\varphi\mid m)$ with parameter $m$, and $K(m)=F(\pi/2\mid m)$ for its complete form. The vertical Bardeen coordinate $\beta_{\rm im}$ is signed so that a backward-traced ray with $\beta_{\rm im}>0$ initially moves toward the pole and turns before reaching the disc. The $\beta_{\rm im}\leq0$ branch reaches the equator directly.

We next solve the radial quartic $\mathcal R(r)=0$ with Ferrari's method. Its real roots identify possible radial turning points, not the emission radius. The exterior four-, two-, and zero-real-root cases use equations (B46), (B75), and (B109) of \citet{GrallaLupsasca2020}. For four real roots we order $r_1<r_2<r_3<r_4$, while for two real roots $r_1<r_2$ are the real pair. Evaluating the observer-connected radial solution with Jacobi functions at $\gamma_d$ from equation~\eqref{eq:app_polar_mino} gives $r_{\rm em}=r(\gamma_d)$. A disc hit requires valid radial and polar domains, an intersection outside the horizon and within the emitting disc, and $g>0$. Higher-order disc crossings are not retained.

At $r_{\rm em}$, we evaluate the disc's circular velocity $u^\mu=u^t(1,0,0,\Omega)$. Normalizing it by $g_{\mu\nu}u^\mu u^\nu=-1$ gives
\begin{equation}
 u^t=\left[-\left(g_{tt}+2\Omega g_{t\phi}
 +\Omega^2g_{\phi\phi}\right)\right]^{-1/2},
\label{eq:circular_disk_ut}
\end{equation}
with $g_{tt}=-(1-2/r)$, $g_{t\phi}=-2a/r$, and $g_{\phi\phi}=r^2+a^2+2a^2/r$ on the equatorial plane \citep{Cunningham73,PageThorne1974}. Substituting $u^t$, $\Omega$, and the photon's conserved $\lambda$ into equation~\eqref{eq:kerr_gfactor} then gives that pixel's frequency shift, without spatial deposition or resampling.

\subsection{Arrival times and emission azimuth}
\label{app:kerr_coordinates}

Having found the disc intersection, we integrate the time and azimuth components of the same geodesic equations along that ray. Following equations (10)--(13) of \citet{GrallaLupsasca2020}, the polar contributions use
\begin{equation}
 G_\theta=\int_{\rm ray}\frac{d\theta}{\pm_\theta\sqrt{\Theta}},
 \qquad\quad G_\phi=\int_{\rm ray}\frac{\csc^2\theta\,d\theta}{\pm_\theta\sqrt{\Theta}},
 \qquad\quad G_t=\int_{\rm ray}\frac{\cos^2\theta\,d\theta}{\pm_\theta\sqrt{\Theta}} .
 \label{eq:app_polar_coordinate_integrals}
\end{equation}
For the observer-to-first-disc path, $G_\theta=\gamma_d$, while $G_\phi$ and $G_t$ supply the polar contributions to azimuth and coordinate time. The signs follow the motion and reverse at turning points. The corresponding radial contributions are
\begin{equation}
 \Phi_r=\int_{\rm ray}\frac{a(2r-a\lambda)}{\Delta}
                  \frac{dr}{\pm_r\sqrt{\mathcal R}},
 \qquad\quad
 T_r=\int_{\rm ray}\left[r^2+\frac{2r(r^2+a^2-a\lambda)}{\Delta}\right]
                  \frac{dr}{\pm_r\sqrt{\mathcal R}} .
 \label{eq:app_radial_coordinates}
\end{equation}
Adding the radial and polar contributions gives the azimuth and travel time. We evaluate the polar integrals and four-real-root radial endpoints through Carlson symmetric integrals $R_F$, $R_D$, and $R_J$ \citep{Carlson_1995,YangWang2013}. Equations (29)--(35) and (B36)--(B44) of \citet{GrallaLupsasca2020} give the corresponding Legendre expressions. The Legendre-to-Carlson identities are given in equations (19.25.5), (19.25.9), and (19.25.14) of \citet{NISTDLMF}, whose modulus $k$ corresponds to our parameter $m=k^2$.

We end the radial integrals at a common finite observer radius $r_o\geq3000$ outside the emitting disc to avoid the divergent travel time to infinity. The polar integral includes the path to infinity, so we subtract its outer time contribution $G_t^{\rm tail}\simeq\gamma_o\cos^2i$, where $\theta\simeq i$ and $\gamma_o=\int_{r_o}^{\infty}dr/\sqrt{\mathcal R(r)}$. We evaluate this outer Mino-time interval by eight-point quadrature in $u=1/r$. With observer azimuth $\phi_o=0$, we store
\begin{equation}
 \Delta\phi=\Phi_r+\lambda G_\phi,\qquad\quad
 \phi_{\rm em}^{\rm BL}=-\Delta\phi,\qquad\quad
 t_{\rm do}=T_r+a^2\left(G_t-\gamma_o\cos^2i\right),
\label{eq:app_coordinate_reconstruction}
\end{equation}
where $\Delta\phi=\phi_o-\phi_{\rm em}^{\rm BL}$ is the source-to-observer azimuthal displacement. Axisymmetric emission does not require the azimuth, but variability requires $t_{\rm do}$. We subtract its minimum valid value, then combine it with the lamp travel time using equation~\eqref{eq:total_disk_delay}. Appendix~\ref{app:kerr_implementation} describes numerical checks and fallback evaluations.

\subsection{Finite-axis lamppost transfer}
\label{app:kerr_lamppost}
We separately trace photons from the lamp to the disc to obtain its heating profile and illumination delay. Their conserved quantities are set by the launch direction, independently of the observer rays. For $h=H_{\rm lamp}/R_g$, rays start in the local orthonormal frame of a static lamp at $(h,\theta_s)$, with $\theta_s=0$ for the upper axial lamp. We sample launch angles $0\leq\alpha<\pi$, measured from the inward radial direction, with $\sin\alpha$ solid-angle weights. Thus $\alpha<\pi/2$ launches inward and $\alpha>\pi/2$ outward. Axial rays have $\lambda=0$, and the local direction fixes the Carter constant,
\begin{equation}
 q^2=
 \frac{\Sigma_s\sin^2\alpha}{-g_{tt,s}}
 -a^2\cos^2\theta_s,\qquad
 \Sigma_s=h^2+a^2\cos^2\theta_s,
\label{eq:app_lamp_carter}
\end{equation}
where $-g_{tt,s}=1-2h/\Sigma_s>0$ is required for a static source. We again solve the polar motion to the equator and evaluate the radial solution at the same parameter interval, now starting at the finite lamp height rather than infinity. The initial radial phase is set by the lamp radius and inward or outward launch direction. We use the infinity-connected radial branch unless $r_2\leq h\leq r_3$ for four real roots, which requires the inner branch in equation (B27) of \citet{GrallaLupsasca2020}. Retaining both possibilities avoids discarding valid rays at low lamp heights and high spins.

The frequency shift is the ratio of photon energies measured in the orbiting disc and static lamp frames \citep[Appendix B]{Ingram2019},
\begin{equation}
 g_{\rm ld}\equiv\frac{\nu_{\rm disc}}{\nu_{\rm lamp}}
 =\frac{\sqrt{-g_{tt,s}}}{g_{\rm d\infty}},
\label{eq:app_lamp_gfactor}
\end{equation}
where $g_{\rm d\infty}=1/u^t$ follows from equation~\eqref{eq:kerr_gfactor} with $\lambda=0$. Thus $g_{\rm ld}=\sqrt{-g_{tt,s}}u^t$ reuses equation~\eqref{eq:circular_disk_ut} at the receiving disc radius, without an additional ray trace.

The same coordinate-time integrals, evaluated from lamp to disc, give the illumination delay,
\begin{equation}
 t_{\rm ld}=T_{r,{\rm ld}}+a^2G_{t,{\rm ld}}.
\label{eq:app_lamp_time}
\end{equation}
The polar integral is analytic, while the one-dimensional radial integral uses 24-point Gauss--Legendre quadrature. The Fig.~\ref{fig:end_to_end_hst_lsst} source uses 1024 launch angles and 512 radial annuli for both the light curves and response functions.

To turn the ray distribution into illumination, we rebin from the ISCO to the outer radius of the finite source model into annuli that conserve the intercepted lamp solid angle. This outer radius is distinct from the normalization limit $R_{\rm out}$ in Appendix~\ref{app:disk_profile}. We then divide by the receiving area and apply the energy and arrival-rate factors in equation~\eqref{eq:gr_lamppost_kernel}. The area measured in the orbiting disc frame is \citep[Appendix A]{Ingram2019}
\begin{equation}
 \frac{dA_{\rm proper}}{R_g^2}
 =2\pi\gamma_\phi(r)
 \sqrt{g_{rr}(r)g_{\phi\phi}(r)}\,dr,
\label{eq:app_proper_annulus}
\end{equation}
with
\begin{equation}
 g_{rr}=\frac{r^2}{\Delta},\qquad\quad
 g_{\phi\phi}=r^2+a^2+\frac{2a^2}{r},\qquad\quad
 \gamma_\phi=\alpha_{\rm Z}u^t,\qquad\quad
 \alpha_{\rm Z}=\sqrt{\frac{\Delta}{g_{\phi\phi}}},
\end{equation}
on the equatorial plane. Here $\alpha_{\rm Z}$ is the lapse of a zero-angular-momentum observer, and $\gamma_\phi$ converts the stationary spatial annulus to the disc-comoving area. Areas are in units of $R_g^2$. For extreme annuli not reached by accepted launch angles, we use the dimensionless flat-space continuation

\begin{equation}
 {\cal K}_{\rm flat}(r)=
 \frac{4h/3}
 {(r^2+h^2)^{3/2}},
\label{eq:app_flat_lamp}
\end{equation}
whose radial dependence is the Newtonian illumination profile \citep{Dauser2013}. The factor $4/3$ combines with the $3/(8\pi)$ prefactor in equation~\eqref{eq:lamp_temperature} to recover the adopted two-lamppost Newtonian flux in equation~\eqref{eq:flat_lamppost_flux}. We match the continuation's amplitude and Euclidean travel time to the nearest sampled relativistic annulus, then shift all delays by one common zero point. Finally, for the resulting profile ${\cal K}^{\rm raw}_{\rm lamp}$, we adopt the normalization
\begin{equation}
 {\cal K}_{\rm lamp}(r)
 ={\cal K}^{\rm raw}_{\rm lamp}(r)
 \frac{\displaystyle\int_{\rm disc}{\cal K}_{\rm flat}(r)\,dA_{\rm proper}}
      {\displaystyle\int_{\rm disc}{\cal K}^{\rm raw}_{\rm lamp}(r)\,dA_{\rm proper}}.
\label{eq:app_lamp_normalization}
\end{equation}
The integrals are sums over annuli weighted by proper area. This prescription retains the relativistic radial redistribution while fixing the integrated heating to the flat-profile value, so its overall amplitude remains controlled by $f_{\rm lamp}$.

\subsection{Thermal emission and response weights}
\label{app:source_observables}

For each observer pixel, we evaluate the lamp heating and illumination delay at its emission radius. Combining lamp-to-disc and disc-to-observer travel times gives the delay in equation~\eqref{eq:total_disk_delay}. The delayed driver and viscous heating then set $T_p$ through equation~\eqref{eq:time_dependent_temperature}. With frequency shift $g_p$ and rest-frame wavelength $\lambda_{b,\infty}$, the colour-corrected intensity is
\begin{equation}
 I_{\nu,\infty,p}=
 \frac{1}{f_{\rm col}^4}
 \frac{2hc}{\lambda_{b,\infty}^{3}}
 \left\{
 \exp\!\left[
 \frac{hc}
 {\lambda_{b,\infty}g_p f_{\rm col} k_{\rm B}T_p}
 \right]-1
 \right\}^{-1}.
\label{eq:relativistic_blackbody}
\end{equation}
Here $h$ and $k_{\rm B}$ are the Planck and Boltzmann constants. This is the invariant transformation $I_{\nu,\infty}=g^3I_{\nu/g,{\rm em}}$ specialized to a diluted blackbody. The factor $f_{\rm col}^{-4}$ preserves the bolometric flux. Cosmological dimming and the pixel solid angle are applied as described in Section~\ref{subsec:observer_brightness}.

The response weight measures how this intensity changes under a small lamp-amplitude perturbation about $X=1$. Writing $x_b=hc/(\lambda_{b,\infty}g f_{\rm col}k_{\rm B}T_{\rm eff})$, differentiating the temperature and Planck functions gives
\begin{equation}
 W_b(\boldsymbol\beta)\equiv
 \left.\frac{\partial I_{\nu,b}^{\rm obs}(\boldsymbol\beta,X)}
 {\partial X}\right|_{X=1}
 =I_{\nu,b}^{\rm obs}(\boldsymbol\beta)
 \frac{x_b(\boldsymbol\beta)e^{x_b(\boldsymbol\beta)}}
 {e^{x_b(\boldsymbol\beta)}-1}
 \frac{T_{\rm lamp}^4(\boldsymbol\beta)}
 {4T_{\rm eff}^4(\boldsymbol\beta)}.
\label{eq:band_response_weight}
\end{equation}
All quantities on the right are evaluated at $X=1$. The Planck factor gives the intensity's temperature sensitivity, while $T_{\rm lamp}^4/(4T_{\rm eff}^4)$ gives the fractional temperature response to the lamp. Weighting pixels by their microlensing magnification and grouping them by delay gives equation~\eqref{eq:microlensed_response}. We distribute each pixel's response between neighbouring delay bins, then average monotone cubic Hermite interpolations of the cumulative response on four shifted grids. Differences between cumulative values at the requested bin edges give the response integrated over each bin. The reconstruction preserves the integrated response and supplies both the displayed transfer functions and the optional response-mode convolutions. Mean lags are evaluated directly from the pixel weights. The non-linear light-curve calculation does not use this reconstruction.

Convolving the response with the driver perturbations and adding the baseline flux gives the linear approximation, with the magnification pattern held fixed over the response delay,
\begin{equation}
 F_{\nu,i,b}^{\,\rm L}(t)\simeq F_{\nu,i,b}^{\,{\rm L},(0)}(t)
 +\int_0^\infty
 \psi_{i,b}(\tau;t)\delta X(t-\Delta t_i-\tau)\,d\tau,
 \label{eq:linear_response_approximation}
\end{equation}
where $\delta X(t)=X(t)-1$ and $F_{\nu,i,b}^{\,{\rm L},(0)}$ is the microlensed baseline flux. Microlensing changes the baseline, response amplitude, and response shape. The package offers linear- and quadratic-response modes for dense time-and-wavelength calculations, but their accuracy is not established by the present benchmarks.

\subsection{Mock image and observing prescription}
\label{app:mock_observations}

The mock HST-like image in Fig.~\ref{fig:end_to_end_hst_lsst} contains a common elliptical power-law plus external-shear lens, extended lens-galaxy light, a lensed S\'ersic host, and the variable quasar point images. It uses a $160^2$ grid with 0.025 arcsec pixels and a 0.08 arcsec full width at half maximum point-spread function, with a realization of Poisson and read noise. The light-curve samples are generated separately from the source fluxes, not extracted from this mock image.

For visit depth $m_5$, noiseless magnitude $m$, and $x_{\rm phot}=10^{0.4(m-m_5)}$, the LSST magnitude uncertainty is
\begin{equation}
\sigma_{\rm rand}^2=(0.04-\gamma_b)x_{\rm phot}+\gamma_b x_{\rm phot}^2,
\qquad
\sigma_m^2=\sigma_{\rm rand}^2+\sigma_{\rm sys}^2.
\label{eq:lsst_photometric_noise}
\end{equation}
We use $\sigma_{\rm sys}=0.005$ mag, $\gamma_u=0.038$, and $\gamma_{g,r,i,z,y}=0.039$ \citep{LSST,Suberlak_2021,Sheng_2022}. These mock survey samples omit an additional image-deblending error model.

\subsection{GPU execution and numerical checks}
\label{app:kerr_implementation}

\begin{figure*}
    \centering
    \includegraphics[width=\linewidth]{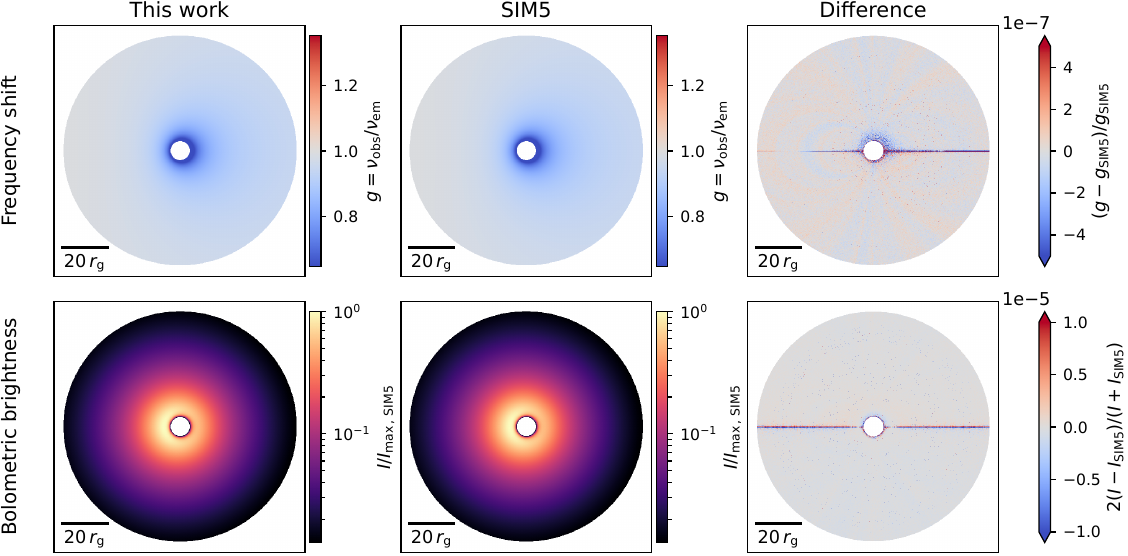}
    \caption{
Visual validation of the direct primary observer-to-disc analytic PyTorch Kerr transfer against SIM5 for the Q2237+0305 fiducial spin and inclination. The same matched calculation is also performed for RX~J1131$-$1231 and HE~0435$-$1223. The panels show the frequency shift, observed brightness, and signed fractional residuals for Q2237+0305. Typical residuals are consistent with float32 numerical precision, with larger differences confined to localized regions. Colourbar extensions mark residuals outside the displayed range. Both calculations reach exactly the same observer-screen pixels in all three cases. Image NMSE ranges from $6.2\times10^{-11}$ to $1.2\times10^{-8}$, and integrated flux differs by at most $3.7\times10^{-7}$.     }
    \figurealttext{Matched PyTorch and SIM5 images show relativistic frequency shifts and disc brightness, followed by signed residual maps. The two calculations reproduce the same spatial structure, with much smaller differences in the residual panels.}
\label{fig:gr_sim5_validation}
\end{figure*}

The solvers use PyTorch tensor operations on CPU or CUDA, with float32 real-pair representations of Ferrari roots, Jacobi functions, and Carlson $R_F$, $R_D$, and $R_J$ algorithms. Fixed iteration counts support \texttt{torch.compile} and CUDA graphs, while spatial chunking bounds memory use. For rays from the axial lamp, we evaluate the polar integral analytically and use a numerically stable expression for the radial potential at launch, avoiding loss of precision for nearly tangential rays.

Roots are classified as real when $|\operatorname{Im}r_j|\leq\sqrt{\epsilon_{\rm mach}}\max(1,|r_j|)$. Numerical checks require the radial integral of $1/\sqrt{\mathcal R}$ to reproduce $\gamma_d-\gamma_o$ and the polar integral $G_\theta$ to reproduce $\gamma_d$. Ill-conditioned polar endpoints are reevaluated in float64 while retaining the radial solution. Rays outside the four-real-root analytic coordinate domain, or failing an endpoint check, use endpoint-regularized Gauss--Legendre quadrature. Any remaining failed coordinates are collected and retried once in float64. A pixel is retained only if its coordinates are finite and the radial and polar solutions agree on the elapsed Mino time.

The SIM5 comparison in Section~\ref{subsec:gr_validation} and Fig.~\ref{fig:gr_sim5_validation} validates the primary observer-to-disc image and transfer, not the lamppost heating or response delays \citep{Sim5,Sim5_2}. We define the image NMSE as $\sum_p (I_p-I_{p,\mathrm{SIM5}})^2/\sum_p I_{p,\mathrm{SIM5}}^2$, where $p$ indexes image pixels. Separate package tests compare lamp intersections with independent geodesic integration, selected lamp travel times with numerical quadrature, and float32 illumination profiles with float64 calculations.

\section{Expanding Supernova Source Example}
\label{app:supernova}

Our prototype defines the explosion epoch in the observer frame and all evolutionary time-scales in the rest frame. Its bolometric luminosity is
\begin{equation}
L(t)=
\begin{cases}
0, & t\leq0,\\
L_{\rm peak}(t/t_{\rm peak})^{2.4},
   & 0<t\leq t_{\rm peak},\\
L_{\rm peak}\exp[-(t-t_{\rm peak})/t_{\rm fall}],
   & t>t_{\rm peak}
\end{cases}.
\label{eq:supernova_luminosity}
\end{equation}
Here $L_{\rm peak}$ is the peak bolometric luminosity, $t_{\rm peak}$ is its rest-frame peak time, and $t_{\rm fall}$ is the post-peak exponential decay time-scale. The adopted parameters are $L_{\rm peak}=10^{43.2}\,\mathrm{erg\,s^{-1}}$, $t_{\rm peak}=18\,$days, and $t_{\rm fall}=28\,$days. The photosphere expands as $R_{\rm ph}=R_0+v_{\rm ph}t$ and its colour temperature follows $L=4\pi R_{\rm ph}^2\sigma_{\rm SB}T^4$. We adopt $R_0=10^{11}\,$cm, $v_{\rm ph}=1.2\times10^4\,\mathrm{km\,s^{-1}}$, and temperature bounds of $3000$--$18000\,$K. Time in the expansion law is converted to seconds. When a temperature bound is active, a wavelength-independent dilution factor renormalizes the bounded blackbody to the prescribed bolometric luminosity. This prevents photospheric expansion from producing an artificial late-time rebrightening. A blackbody continuum and a phenomenological rest-UV blanketing factor supply six-band $F_\nu$ images. The projected ellipse, position angle, and linear limb-darkening profile evolve on the fixed source grid. Every band-dependent spatial correction is renormalized independently, so it redistributes surface brightness without changing the corresponding band flux. The adopted photospheric parameters are representative prototype values. The approximately three-week achromatic phase is typical rather than universal, but variation in microlensing configuration and filter choice can shorten it to only a few days \citep{Huber_2021}. We restrict interpretation to the photospheric phase and do not use this source as a distance estimator or substitute for radiative-transfer templates such as those considered by \citet{Huber_2021}. The matched 12-macroimage transient benchmark uses 51 map epochs over 200 observer-frame days at a four-day cadence. It uses the same frame-zero dense $k=1$ scout to normalize the subsequent $k=2$ selected maps as in the adopted fast quasar configuration. The correction is source-model independent.

\begin{figure*}
    \centering
    \includegraphics[width=\linewidth]{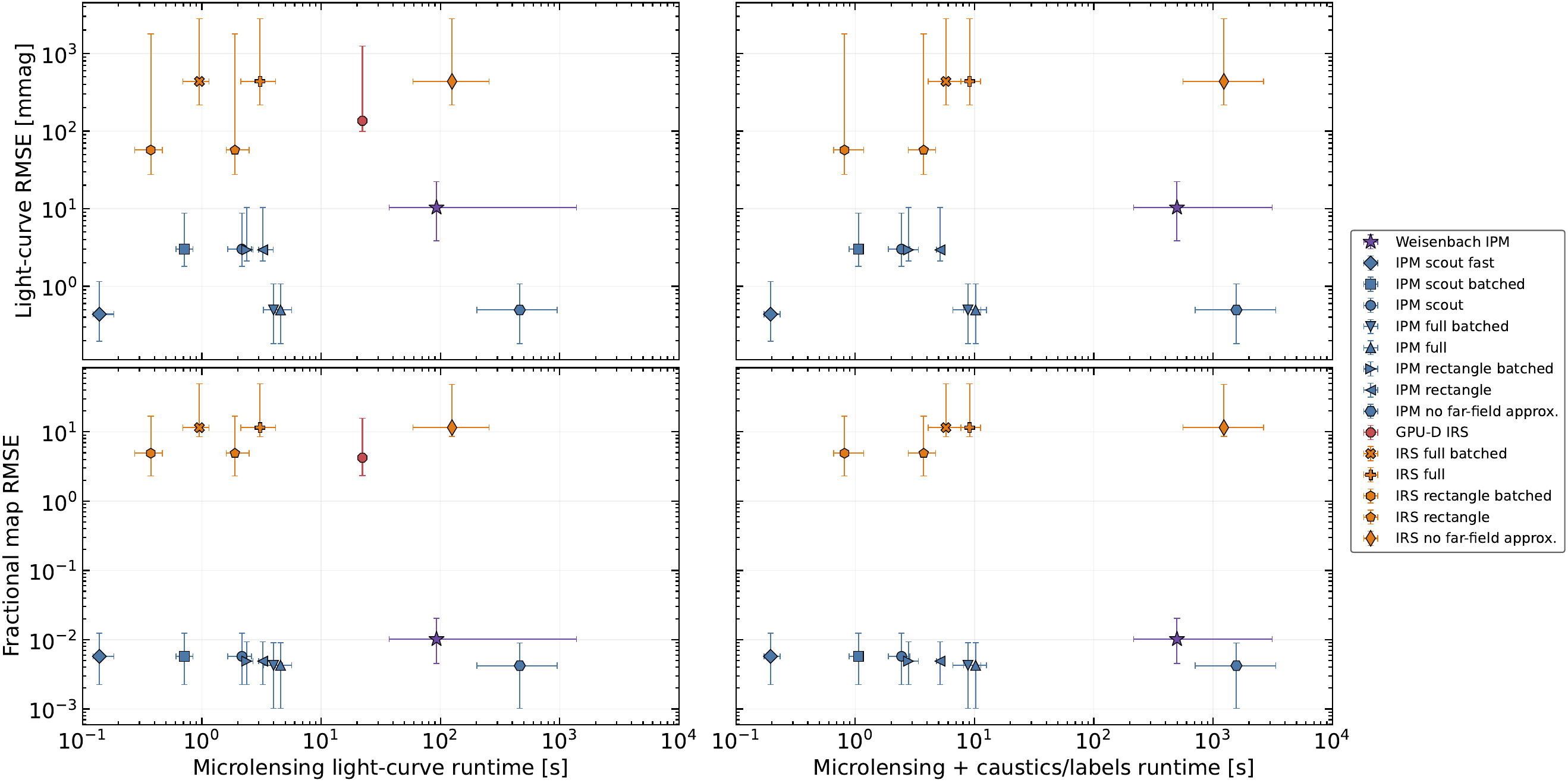}
    \caption{
Accuracy--runtime comparison for the evolving supernova source across the same 12 macroimages as the quasar benchmark. Internal light-curve runtimes refer to all 51 dynamical-map epochs, and fractional map RMSE uses every pixel at every epoch. Light-curve metrics omit only samples below 1 per cent of the peak reference flux in that observer band. For the few above-threshold GPU-D samples with zero candidate flux, the magnitude residual uses the 1 per cent flux-ratio floor. Weisenbach light-curve timing and accuracy use all 51 epochs. Its map errors compare the complete native square with a reference evaluated directly on the same grid. Caustic-stage timing follows Section~\ref{subsec:benchmark_contract}. GPU-D IRS appears only in the left panels because it does not calculate caustics or labels. Its plotted runtime is the complete measured 51-frame executable process time.
    }
    \figurealttext{Four log-scale panels compare supernova light-curve and map errors against runtime, without and with caustic labels. Different symbols identify the numerical methods. The source-scouted calculations achieve low light-curve errors at short runtimes.}
\label{fig:supernova_accuracy_runtime}
\end{figure*}

Fig.~\ref{fig:supernova_accuracy_runtime} summarizes the corresponding accuracy and runtime distribution across the 12 macroimages. The fast source-scouted biquadratic IPM method has a median light-curve RMSE of $0.44^{+0.72}_{-0.24}\,$mmag and a fractional map RMSE of $5.8^{+6.7}_{-3.5}\times10^{-3}$. The complete light curve takes $0.14^{+0.044}_{-0.016}\,$s, or $0.20^{+0.039}_{-0.023}\,$s with caustics and labels. The Weisenbach comparison evaluates all 51 epochs, with median light-curve and fractional map RMSEs of $10\,$mmag and $1.0\times10^{-2}$. Its complete measured light-curve runtime is $93\,$s, increasing to $500\,$s when the extrapolated caustic cost is included. Across individual macroimages, the fast IPM scout method has a median single-curve speed advantage of $520$ for the light curve and $2.5\times10^3$ when the Weisenbach caustic calculation is included. Batching up to four curves increases these median advantages to $620$ and $3.6\times10^3$.

We separately measured throughput for the 51-epoch supernova sequence, which spans 0--200 observer-frame days at a four-day cadence. Batching up to four separate light-curve calculations with matched seeded inputs decreases the amortized median runtime from $0.14^{+0.044}_{-0.016}\,$s to $0.12^{+0.041}_{-0.015}\,$s per curve. The median throughput gain is $1.17^{+0.064}_{-0.062}$ times. For the workload that includes caustics and source-centre labels, the same batching lowers the amortized per-curve runtime from $0.20^{+0.039}_{-0.023}\,$s to $0.13^{+0.038}_{-0.016}\,$s per curve. The corresponding median throughput gain is $1.38^{+0.13}_{-0.087}$ times.

We also evaluate the published GPU-D implementation for all 51 epochs in each macroimage. At the same nominal $10^7$-ray budget, its median complete process time is $22^{+1.7}_{-1.2}\,$s for the map sequence. Across matched macroimages, this is $160^{+25}_{-24}$ times slower than the complete fast IPM scout light curve, even though GPU-D does not integrate the source brightness over the magnification maps or calculate caustics and labels. The much smaller transient field is severely undersampled by direct ray deposition at this budget. The median raw fractional map RMSE is $4.2$ and 93 per cent of its source-plane pixels receive no ray. The 12 macroimage parameter sets deliberately match the quasar benchmark and are therefore a numerical stress test, not an observability-selected lensed-supernova sample. 

\section{Analytic point-lens convergence}
\label{app:point_lens_convergence}

\begin{figure}
    \centering
    \includegraphics[width=0.5\textwidth]{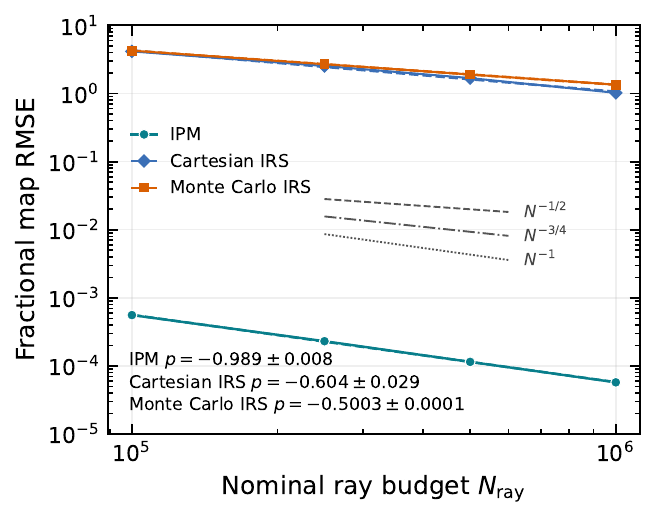}
    \caption{
Analytic point-lens convergence over $10^5$--$10^6$ nominal rays. Teal circles show the fiducial $r=2$, $v=4$ biquadratic IPM rule. Blue diamonds show Cartesian IRS. Orange squares and their shaded 16th--84th percentile envelope show eight independent Monte Carlo IRS realizations. Dashed coloured lines are log--log fits over the complete displayed interval.}
    \figurealttext{A log-scale plot compares point-lens map errors for IPM, Cartesian IRS, and Monte Carlo IRS as the ray count increases. Dashed fits show the convergence rates, and shading shows the spread among independent Monte Carlo realisations.}
\label{fig:point_lens_convergence}
\end{figure}

An isolated point lens provides an exact end-to-end convergence test without the finite-reference uncertainty of a many-star magnification map. In units of its Einstein radius, the point-source magnification is
\begin{equation}
 \mu_{\rm pt}(u)=\frac{u^2+2}{u\sqrt{u^2+4}},
\label{eq:point_lens_magnification}
\end{equation}
where $u$ is the angular lens--source separation in units of the point-lens Einstein radius. We compare the numerical maps with this analytic solution using the fractional map RMSE defined in equation~\eqref{eq:map_fractional_rmse}. The $1024^2$ source grid spans $4\theta_{\rm E}$ on a side, the lens grid spans $6\theta_{\rm E}$, and the reported norm uses $0.08\leq u\leq1.8$ to exclude the singular point caustic and the finite-field boundary. The IPM calculation retains the fiducial $k=2$, $r=2$, $v=4$ scout, scalar-correction, and biquadratic-polygon settings, while evaluating the sole point lens directly so that an unnecessary far-field approximation cannot set the floor.

Fig.~\ref{fig:point_lens_convergence} fits four ray budgets from $10^5$ to $10^6$. Monte Carlo IRS, shown as the median of eight independent realizations, gives $p=-0.5003\pm0.0001$ for $\epsilon_\mu\propto N_{\rm ray}^{p}$, recovering the expected $N^{-1/2}$ sampling law. Cartesian IRS gives $p=-0.604\pm0.029$ over the same low-occupancy interval. It converges more rapidly than random sampling but has not yet reached the nominal $N^{-3/4}$ boundary-dominated regime. The source-convolved macroimage experiment in Section~\ref{subsec:map_convergence_rate} suppresses the pixel-occupancy noise and gives the steeper Cartesian IRS rate $p=-0.711^{+0.014}_{-0.036}$. The adopted biquadratic IPM calculation gives $p=-0.989\pm0.008$, recovering the expected $N^{-1}$ piecewise-affine IPM benchmark.

\bsp
\label{lastpage}
\end{document}